\documentclass[11pt]{article}
\newif\iffigs\figstrue

\usepackage{epsfig,latexsym}
\usepackage{amsmath}
\usepackage{verbatim}
\usepackage{mathrsfs}
\usepackage{amssymb}

\DeclareMathAlphabet{\mathpzc}{OT1}{pzc}{m}{it}

 \csname
@addtoreset\endcsname{equation}{section}

\def\gz0{\gamma^{0}}

\def\ba{\begin{array}}
\def\ea{\end{array}}
\def\bec{\begin{center}}
\def\ec{\end{center}}
\def\ba{\begin{align}}
\def\ena{\end{align}}

\def\12{\frac{1}{2}}

\newcommand{\I}{\mathrm{Im}}

\def\beq{\begin{equation}}
\newcommand{\eeq}[1]{\label{#1}\end{equation}}
\def\bea{\begin{eqnarray}}
\newcommand{\eea}[1]{\label{#1}\end{eqnarray}}

\begin{document}

\begin{flushright}
CERN-PH-TH/2015-052\\
\end{flushright}

\vspace{15pt}

\begin{center}


{\Large\sc A String--Inspired Model for the Low--$\ell$ CMB}\\


\vspace{38pt}
{\sc N.~Kitazawa${}^{\; a}$ and A.~Sagnotti${}^{\; b,\,c}$}\\[20pt]

{${}^a$\sl\small Department of Physics, Tokyo Metropolitan University \\ Hachioji, Tokyo 192-0397, JAPAN \\ }
e-mail: {\small \it kitazawa@phys.se.tmu.ac.jp}
\vspace{8pt}

{${}^b$\sl\small Department of Physics, CERN Theory Division\\
CH - 1211 Geneva 23, SWITZERLAND \\ }
\vspace{8pt}

{${}^c$\sl\small
Scuola Normale Superiore and INFN\\
Piazza dei Cavalieri \ 7\\I-56126 Pisa \ ITALY \\}
e-mail: {\small \it sagnotti@sns.it}

\vspace{24pt} {\sc\large Abstract}\end{center}

\setcounter{page}{1}
\noindent We present a semi--analytic exploration of some low--$\ell$ angular power spectra inspired by ``Brane Supersymmetry Breaking''. This mechanism splits Bose and Fermi excitations in String Theory, leaving behind \emph{an exponential potential that is just too steep for the inflaton to emerge from the initial singularity while descending it}. As a result, the scalar generically bounces against the exponential wall, which typically introduces \emph{an infrared depression and a pre--inflationary peak} in the power spectrum of scalar perturbations. We elaborate on a possible link between this phenomenon and the low--$\ell$ CMB. For the first 32 multipoles, combining the hard exponential with a milder one leading to $n_s\simeq 0.96$ and with a small gaussian bump we have attained a reduction of $\chi^{\,2}$ to about 46\% of the standard $\Lambda$CDM setting, with both WMAP9 and PLANCK 2013 data. This result corresponds to a $\chi^{\,2}/DOF$ of about 0.45, to be compared with a $\Lambda$CDM value of about 0.85. The preferred choices combine naturally quadrupole depression, a first peak around $\ell=5$ and a wide minimum around $\ell=20$. We have also gathered some evidence that similar spectra emerge if the hard exponential is combined with more realistic models of inflation. A problem of the preferred examples is their slow convergence to an almost scale--invariant profile.

\vfill

\pagebreak

\newpage
\section{\sc  Introduction}\label{sec:intro}

String Theory (for reviews see \cite{strings1}-\cite{strings5}) is an enticing framework for High--Energy Physics beyond the Standard Model, whose foundations remain however somewhat mysterious. Its supersymmetric versions have received a lot of attention during the last decades because gravity emerges in their stable vacua inevitably and without ultraviolet singularities. Moreover, the low--energy Supergravity \cite{SUGRA1,SUGRA2} (for a recent review see \cite{SUGRA3}) provides precise tools to analyze the theory below the scale $M_s \equiv 1/\sqrt{\alpha'}$ of its massive excitations, with intriguing clues on the nature of dark matter and on the origin of flavor. All current implementations of broken Supersymmetry, however, result in vacuum energy contributions that make it difficult to relate them directly to Particle Physics.

The string scale $M_s$ is typically of the order of the Planck scale, and as a result collider experiments or other high--precision measurements capable of capturing some evidence for String Theory appear difficult to conceive. On the other hand, recent observations of the Universe at large scales are providing sizable, if indirect, evidence for phenomena that occurred during its very early stages, when typical energies were not far below the Planck scale. These types of observations can thus potentially provide some missing clues, and conversely String Theory can perhaps provide important information on the underlying phenomena. In particular, the vacuum corrections that accompany a high--scale breaking of Supersymmetry might have played an interesting role in Cosmology.
\begin{figure}[ht]
\begin{center}$
\begin{array}{cc}
\epsfig{file=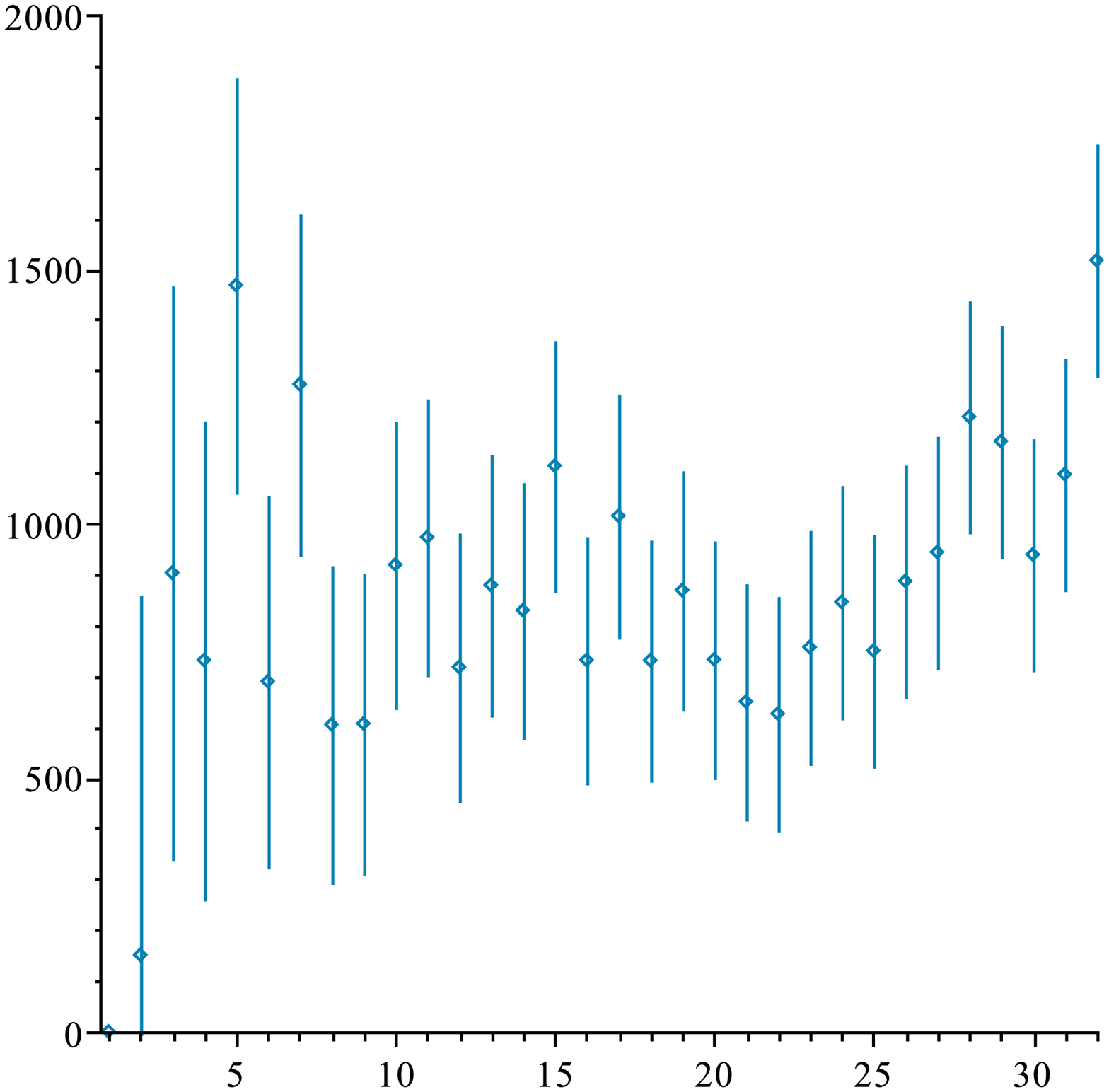, height=1.4in, width=1.4in} & \qquad\quad
\epsfig{file=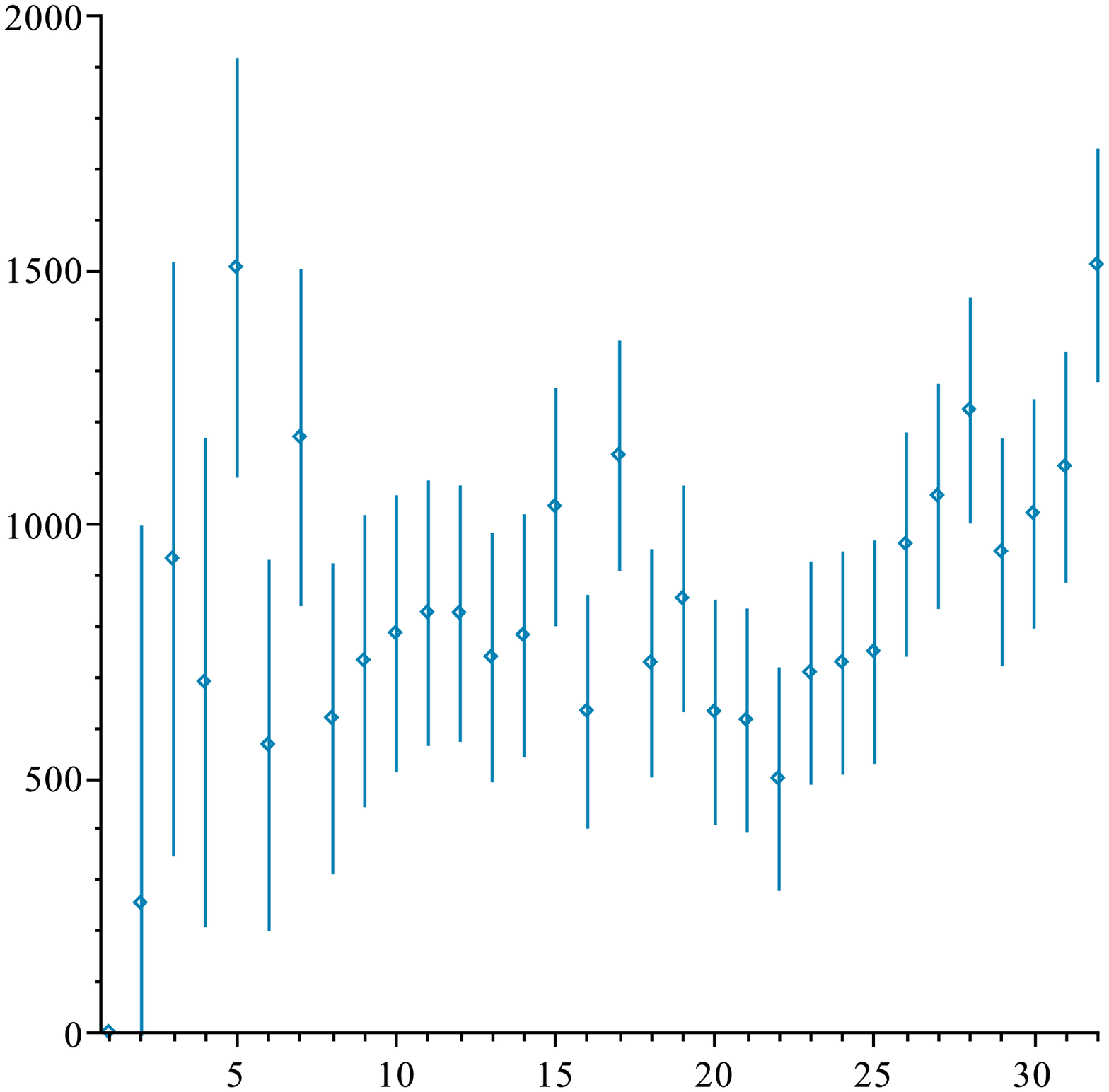, height=1.4in, width=1.4in}
\end{array}$
\end{center}
\caption{\small
The angular power spectra of large--scale CMB scalar perturbations ($\ell \leq 32$)
 observed by WMAP9 (left) and PLANCK 2013 (right).
}
\label{fig:raw_data}
\end{figure}

The temperature anisotropies observed by WMAP \cite{wmap91,wmap92} and PLANCK \cite{PLANCK1}-\cite{PLANCK3} are confronting us with inflaton scalar perturbations consistent with primordial power spectra that are almost scale invariant over about three orders of magnitude, with $P(k) \sim k^{\,n_s-1}$ and $n_s \simeq 0.96$. Following \cite{cm}, these types of spectra are naturally ascribed to quantum fluctuations of a slowly--rolling scalar that were stretched to super--horizon scales by a primordial inflation \cite{inflation1} - \cite{inflation7} (for reviews see \cite{inflation8} - \cite{inflation13}). Remarkably, this is the same phenomenon that is usually invoked to account for the striking flatness and homogeneity of our Universe at large scales. The WMAP and PLANCK experiments have extracted from the resulting framework, usually referred to as $\Lambda$CDM, indications on the matter and energy content of the Universe to unprecedented levels of accuracy.

The low--$\ell$ angular power spectra revealed by WMAP9 and PLANCK 2013 are displayed in fig.~\ref{fig:raw_data}. They are qualitatively very similar and slightly at variance with the almost horizontal profile that $\Lambda$CDM would predict in the region. This discrepancy is widely known and would probably spell trouble for the standard picture, were it not for ``cosmic variance''. Only few independent observations contribute in fact to low multipoles, and this adds more than a word of caution to any attempts to take at face value these facts, which could be mere fluctuations. Keeping this proviso well in mind, we shall try nonetheless to explore further, within the usual one--field description and following \cite{dkps}-\cite{ks2}, the possibility that the discrepancies be real. This is the main standpoint of this work, and the reader should be well aware of it. Moreover, we are pushing indications from the low--energy Supergravity to early epochs, to an extent that is not fully justified \cite{cd}. On the other hand, Supergravity is the only effective tool at one's disposal, while some enticing features of the resulting dynamics appear, to some extent, independent of details that are largely out of control in a top--down approach. These points were stressed clearly in \cite{dks}, \cite{dkps}-\cite{ks2}, but we deemed appropriate to state them anew. Let us also stress that our current analysis rests on simplifications that are only available for low--$\ell$ and confine our semi--analytic tools to the first 30 multipoles or so. This can clearly enhance some effects that we shall display, but we shall soon complement these results with an ongoing full--fledged likelihood analysis \cite{gkmns}.

Some dynamical facts that we shall touch upon generalize the considerations presented in \cite{ks1} and should be of interest in their own right. They concern the behavior of a minimally coupled scalar field in the neighborhood of a ``critical'' exponential wall. One can explore them semi--analytically, since the measured CMB multipoles find a direct counterpart, for low values of $\ell$, in the primordial power spectrum. Peaks and troughs present in the latter translate indeed into similar features of the former \cite{cmbslow}, via a Bessel--like transform. Our model suggests that, \emph{if the discrepancies were real and were captured by the one--field setting, the CMB could be confronting us with some imprints of the onset of inflation.} Needless to say, this scenario would be stunningly exciting. Just as the almost scale invariant scalar perturbations confirmed to high accuracy by PLANCK \cite{PLANCK1}-\cite{PLANCK3} find a rationale in slow--roll motion, so quadrupole depression and other features present in fig.~\ref{fig:raw_data} for low $\ell$ could be telling us something about how this regime was attained. The main difficulty, to which we shall return later, is however that confining the departure from $\Lambda$CDM to one decade or so, as the low--$\ell$ anomalies would require, appears complicated, and more so in models encompassing the exit from inflation.

This letter is devoted to a class of scenarios \cite{dks}, \cite{dkps}-\cite{ks2} that are motivated by ``Brane Supersymmetry Breaking'' (BSB) \cite{bsb1}-\cite{bsb5}. This phenomenon occurs in classically stable vacua of String Theory \cite{orientifolds1} - \cite{orientifolds8} (for reviews see \cite{orientifolds9,orientifolds10}) where branes and orientifolds are individually BPS and yet are collectively incompatible with Supersymmetry. In ten dimensions BSB breaks Supersymmetry at the scale of string excitations, bringing along an exponential potential that is \emph{``critical'', or just steep enough to make it impossible, for the dilaton, to emerge from the initial singularity while descending it} \cite{dks}. For this reason we coined the name ``climbing scalar'' for a field that is generically led to bounce against the exponential wall. Moreover, in \cite{as13,fss} we also argued that the climbing phenomenon persists for $D<10$, since a mixing between the ten--dimensional dilaton and the breathing mode retains a ``critical'' exponential potential even after a compactification.

For the convenient non--canonical scalar $\varphi$ of \cite{dks}, related to a
canonically normalized scalar in $D$ dimensions according to
\beq
 \varphi \ = \ \phi \ \sqrt{\frac{D-1}{D-2}} \ ,
\eeq{phi_varphi}
the ``critical'' potential would read
\beq
V \ = \ T \ e^{ \, 2\, \varphi } \ .
\eeq{onexp}
In \cite{dks} we also stressed that the four--dimensional KKLT uplift \cite{KKLT1,KKLT2} leads precisely to a ``critical'' exponential, and we also showed that the axion is frozen close to the initial singularity. As a result, the ``climbing phenomenon'' continues to occur in that setting, which recently found its place in Supergravity \cite{fkl1} - \cite{fkl3} via the four--dimensional non--linear Supersymmetry introduced, in this context, in \cite{adfs}. Non--linear realizations, however, were associated long ago in \cite{dm_101,dm_102} to BSB in the ten--dimensional Sugimoto model of \cite{bsb1}.

It is natural to explore further some scenarios where the scalar climbs up, bouncing against the exponential wall, as we did in \cite{dkps} and, more recently, in \cite{ks1,ks2}. Even in the absence of concrete top--down information about other regions of the potential, one can indeed try to translate some generic consequences of the bounce into corresponding features of the low--$\ell$ CMB. In \cite{dkps} we elaborated in detail on how an early fast--roll would result in quadrupole depression, if the largest wavelengths that we are observing today exited the horizon when inflation was starting. Moreover the refined analysis in \cite{ks1} showed that \emph{the bounce brings along, in general, a pre--inflationary peak that can lie well apart from the region where the attractor behavior is reached}. This generalizes the feature that accompanies standard transitions to slow--roll nicely described in \cite{destri1,destri2}. Angular power spectra with these pre--inflationary peaks led in \cite{ks1} to an encouraging reduction of $\chi^2/DOF$ for WMAP9 data from 0.855, the value corresponding to $\Lambda$CDM, to about 0.72, even at the cost of adjusting the additional parameter $\varphi_0$.

In four dimensions, the class of potentials
\beq
V(\phi) \ = \ V_0 \ \bigg[ e^{\,\sqrt{6} \, \phi} \ + \ e^{\,\sqrt{6} \, \gamma\, \phi} \bigg] \ = \ V_0 \ \bigg[ e^{\,2 \, \varphi} \ + \ e^{\,2 \, \gamma\, \varphi} \bigg]
\eeq{2exp}
can combine a bounce and an inflationary phase with spectral index $n_s \simeq 0.96$ if $\gamma \simeq \frac{1}{12}$. These toy models are relatively simple \footnote{This potential never vanishes, which allows the convenient gauge choice in \cite{ks1}. The resulting equations are not stiff and can even be studied semi--analytically. Amusingly, one can argue that the branes of String Theory give rise to values of $\gamma$ that are multiples of $\frac{1}{12}$ \cite{as13,fss}.} and were studied in detail in \cite{ks1,ks2}. Moreover, the resulting power spectra of scalar perturbations with pre--inflationary peaks are qualitatively similar to those obtained if the mild exponential in eq.~\eqref{2exp} is replaced with a more realistic Starobinsky potential. This result is consistent with the intuitive idea that bounces give rise to \emph{universal effects that originate in the vicinity of the exponential wall}.

Notice that the initial speed is traded, in this type of dynamics, for the time of the initial singularity. A single integration constant is thus left for $\varphi(t)$, which we shall refer to as $\varphi_0$, as in \cite{dkps,as13,ks1,ks2}; it will be very important in the ensuing discussion, since it determines to which extent the scalar feels the presence of the exponential wall. The non--canonically normalized $\varphi$ of eq.~\eqref{phi_varphi} is a convenient choice in numerical computations, since for one matter it associates the critical value to a fixed exponent for all $D$ \footnote{We would like to stress that, in some of our preceding work, $\phi$ inherited a slightly non--canonical normalization from the string--frame dilaton. This choice introduced an additional factor $\sqrt{2}$ in eq.~\eqref{phi_varphi}.}.

In \cite{ks2} we recently started to explore a modification of eq.~\eqref{2exp},
\beq
V(\phi) \ = \ V_0 \ \left[ e^{\,2 \, \varphi} \ + \ e^{\,2 \, \gamma\, \varphi} \ +  a_1 \, e^{\,-\,a_2\left( \varphi\,+\,a_3\right)^2}\right]\ .
\eeq{2expgauss}
With a positive amplitude $a_1$ of a few percent and with the gaussian bump located close to the exponential wall, this potential can recover, to begin with, the effects in \cite{ks1}, since it simulates an over-critical exponential as the scalar comes close to it. Moreover, it can give rise to an interesting new type of feature, a pair of nearby pre--inflationary peaks followed by a relatively deep trough and a steep rise. This occurs, within a range for $\varphi_0$ and the $a_i$ in eq.~\eqref{2expgauss}, if the scalar climbs up relatively fast, overcomes the bump and then reverts its motion when it is not quite in slow--roll, after being reflected by the exponential wall. We started to play with the bump to obtain larger pre--inflationary peaks, which emerge indeed as $\varphi_0$ is adjusted to that the scalar lingers on it, but we soon realized that this type of behavior can translate into angular power spectra that follow naturally the mean profiles of fig.~\ref{fig:raw_data}. And indeed the very emergence of double peaks led to a collapse of $\chi^2$ to about 12! One might object that three more parameters are introduced via eq.~\eqref{2expgauss}, but the improvement remains sizable even taking this rightfully into account, as can be seen working in terms of the reduced $\chi^{\,2}$. The relevant numbers are in fact $21/29 \simeq 0.72$ for the potential of eq.~\eqref{2exp}, $12/26 \simeq 0.46$ for the potential of eq.~\eqref{2expgauss} and $25.5/30 \simeq 0.85$ for $\Lambda$CDM. While none of the options is excluded, the model with gaussian bump appears clearly preferred.

In trying to connect models inspired by BSB to the CMB, one is clearly forced to complement the key datum drawn from String Theory, the critical exponential that forces an early climbing phase, with other contributions that cannot be justified in a top-down approach, and here we are focussing on two terms. The first is a mild exponential, which grants as in \cite{dkps,ks1,ks2} a spectral index $n_s \sim 0.96$ in the fully developed inflationary phase. The second is a small gaussian bump, a prototype defect. The resulting potentials of eq.~\eqref{2expgauss} are an interesting theoretical laboratory since, depending on the choice of $\varphi_0$, the resulting power spectra can recover the feature described in \cite{destri1,destri2}, their generalizations described in \cite{dkps,ks1,ks2} or new types of double peaks. Moreover, all these effects arise from regions that are close the exponential wall, right before the scalar attains slow--roll, so that the available options appear somehow exhausted in this fashion.

We can now turn to summarize the results of the analysis of angular power spectra in the low--$\ell$ region based on the potentials \eqref{2expgauss}, which was carried out with both WMAP9 and PLANCK 2013 data~\footnote{The PLANCK 2013 data were kindly provided to us by A.~Gruppuso and P.~Natoli, whom we would like to thank also for an ongoing collaboration.}. The present work is meant to complement the analysis in \cite{ks1,ks2} with a scrutiny of new effects introduced by the gaussian bump. For brevity, however, we shall leave aside all technical details about the computation, since they were clearly spelled out in \cite{ks1,ks2}.

\vskip 12pt

\section{\sc  Comparison with the low--$\ell$ CMB}\label{sec:observables}
%
\begin{figure}[ht]
\begin{center}$
\begin{array}{ccc}
\epsfig{file=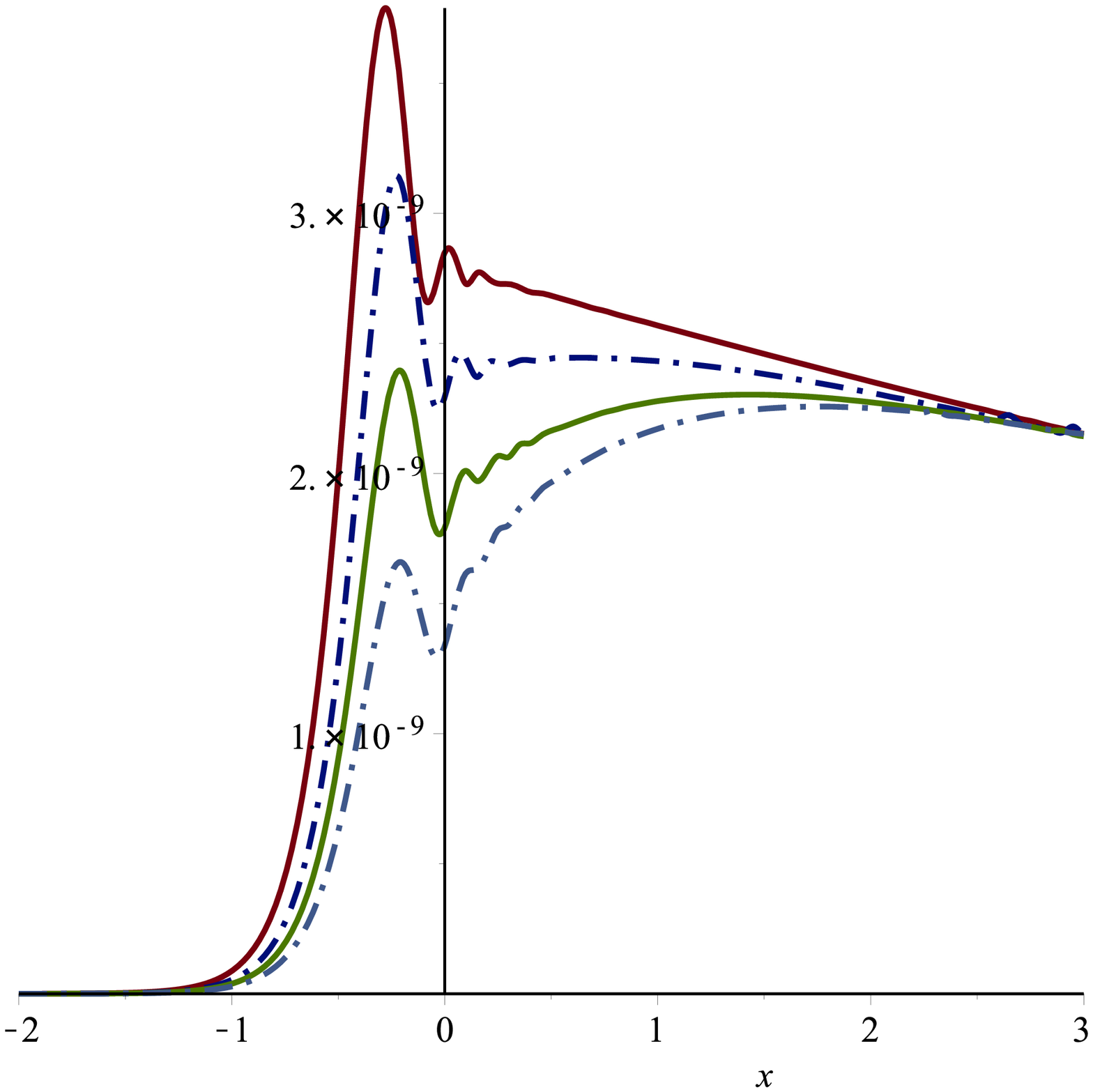, height=1.4in, width=1.4in}& \qquad\quad
\epsfig{file=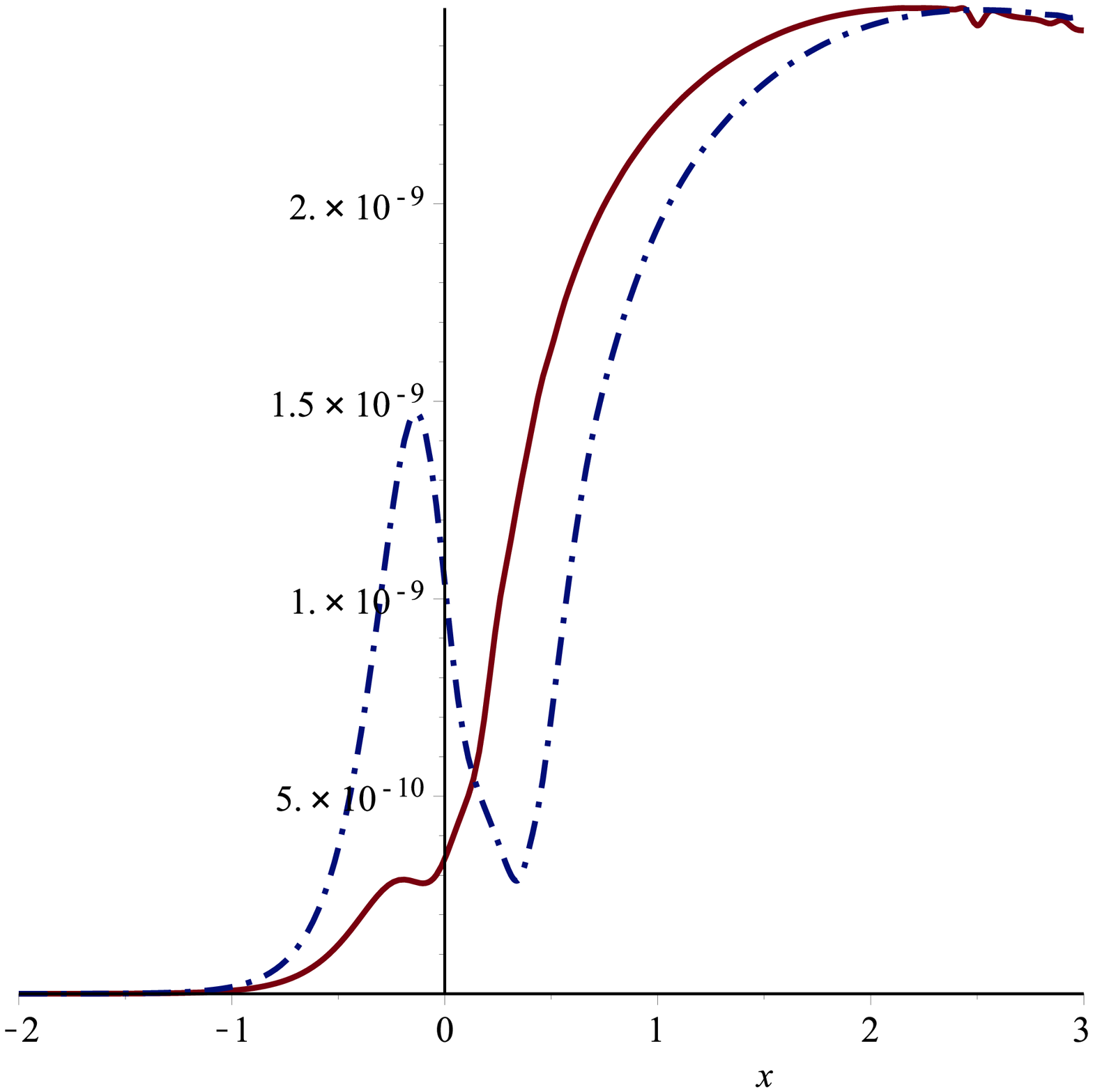, height=1.4in, width=1.4in}& \qquad\quad
\epsfig{file=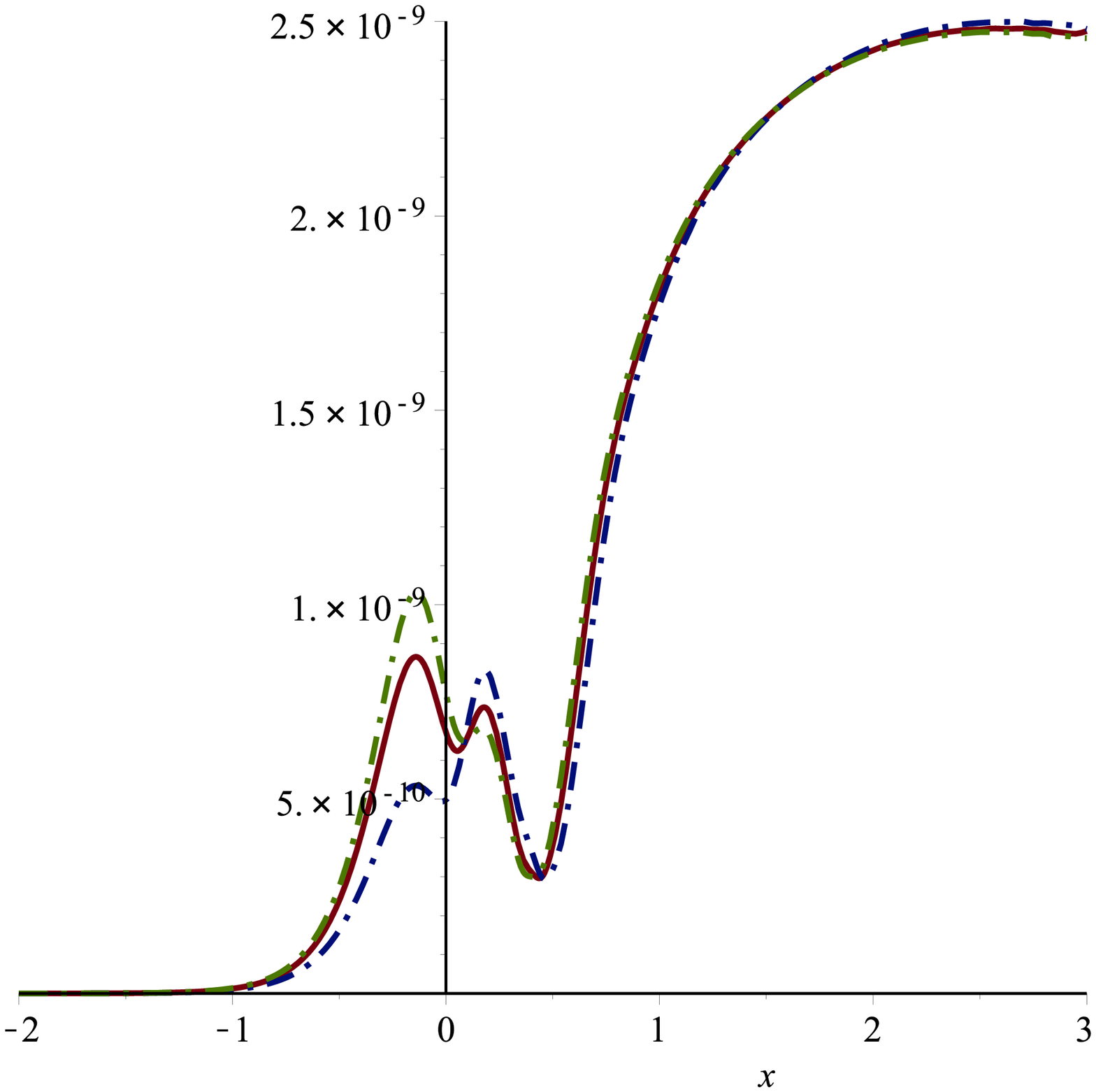, height=1.4in, width=1.4in}
\end{array}$
\end{center}
\caption{\small
Power spectra obtained from the potential of eq.~\eqref{2expgauss} with $(a_1,a_2,a_3)=(0.065,5,1.01)$. The left plot collects some results for the single--peak region, obtained for $\varphi_0=-1.75,-2,-2.5,-4$ (cyan, green, blue, red, or from lower to higher). The middle plot illustrates the central transition region with the two cases $\varphi_0=-0.5,-1$ (blue, red, or from higher to lower). Finally, the right plot collects some results for the double--peak region, obtained for $\varphi_0=-0.3,-0.37,-0.4$ (blue, red, green, or from lower to higher first peak). The first two groups of plots follow the pattern already seen in \cite{ks1,ks2} for the two--exponential potential.}
\label{fig:powerspectra}
\end{figure}

For small values of $\ell$, the relation between primordial power spectrum and CMB angular power spectrum is well captured by \cite{cmbslow}
\beq
A_\ell\;(\varphi_0,{\cal M},\delta) \ = \ {\cal M} \ \ell(\ell+1)\ \int_0^\infty \frac{dk}{k} \ {\cal P}_\zeta \big( k , \varphi_0 \big) \, {j_\ell}^2 \big( k \, 10^\delta \big) \, ,
\eeq{bessel}
where we have emphasized the dependence on $\varphi_0$. Here $j_\ell$ denotes a spherical Bessel function, and nicely enough late time effects are negligible for large scales, or small $\ell \lesssim 35$. The reader should notice the two parameters, ${\cal M}$ and $\delta$, which appear in eq.~\eqref{bessel}. The overall normalization ${\cal M}$ combines various constants that enter the link between angular power spectrum and primordial power spectrum of scalar perturbations, and ultimately reflects the Hubble scale during inflation. On the other hand, $\delta$ controls the horizontal displacement of features introduced by the power spectra for a given $\ell$. In more physical terms, $\delta$ allows a tuning between the wavelengths that are now entering the cosmic horizon and those that were exiting in the pre--inflationary era that we are focussing on.

Our comparisons with low--$\ell$ CMB data rest on the simplest estimator,
\beq
\chi^2 \ = \ \ \sum_{\ell=2}^{32} \frac{\left(A_\ell\;(\varphi_0,{\cal M},\delta)\ - \ A_\ell^{\rm OBS}\right)^2}{\left(\Delta A_\ell^{\rm OBS}\right)^2} \ ,
\eeq{chi_squared}
where $A_\ell^{\rm OBS}$ are the observed central values and $\Delta A_\ell^{\rm OBS}$ are corresponding (symmetric) errors. For the class of potentials of eq.~\eqref{2expgauss} we have performed detailed investigations using data provided by the WMAP and PLANCK collaborations. For various sets of parameters $(a_1,a_2,a_3)$ of the gaussian bump we thus explored a sequence of about 25 values of $\varphi_0$ including $0,-0.25,-0.5,,\ldots,-3.75,-4$, which suffice to capture the evolution of different types of pre--inflationary peaks.
\begin{figure}[ht]
\begin{center}$
\begin{array}{ccc}
\epsfig{file=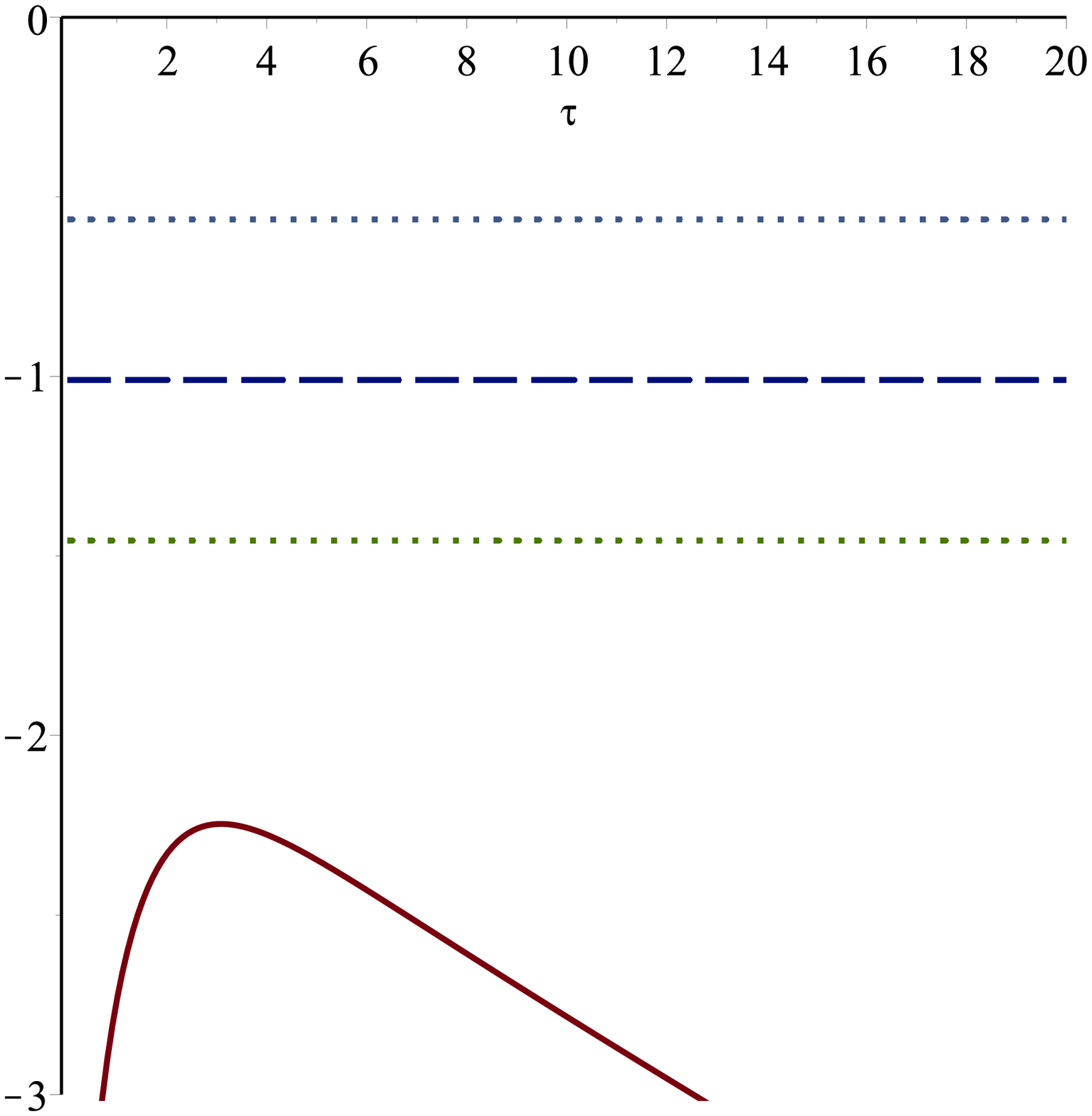, height=1.3in, width=1.3in}& \qquad\quad
\epsfig{file=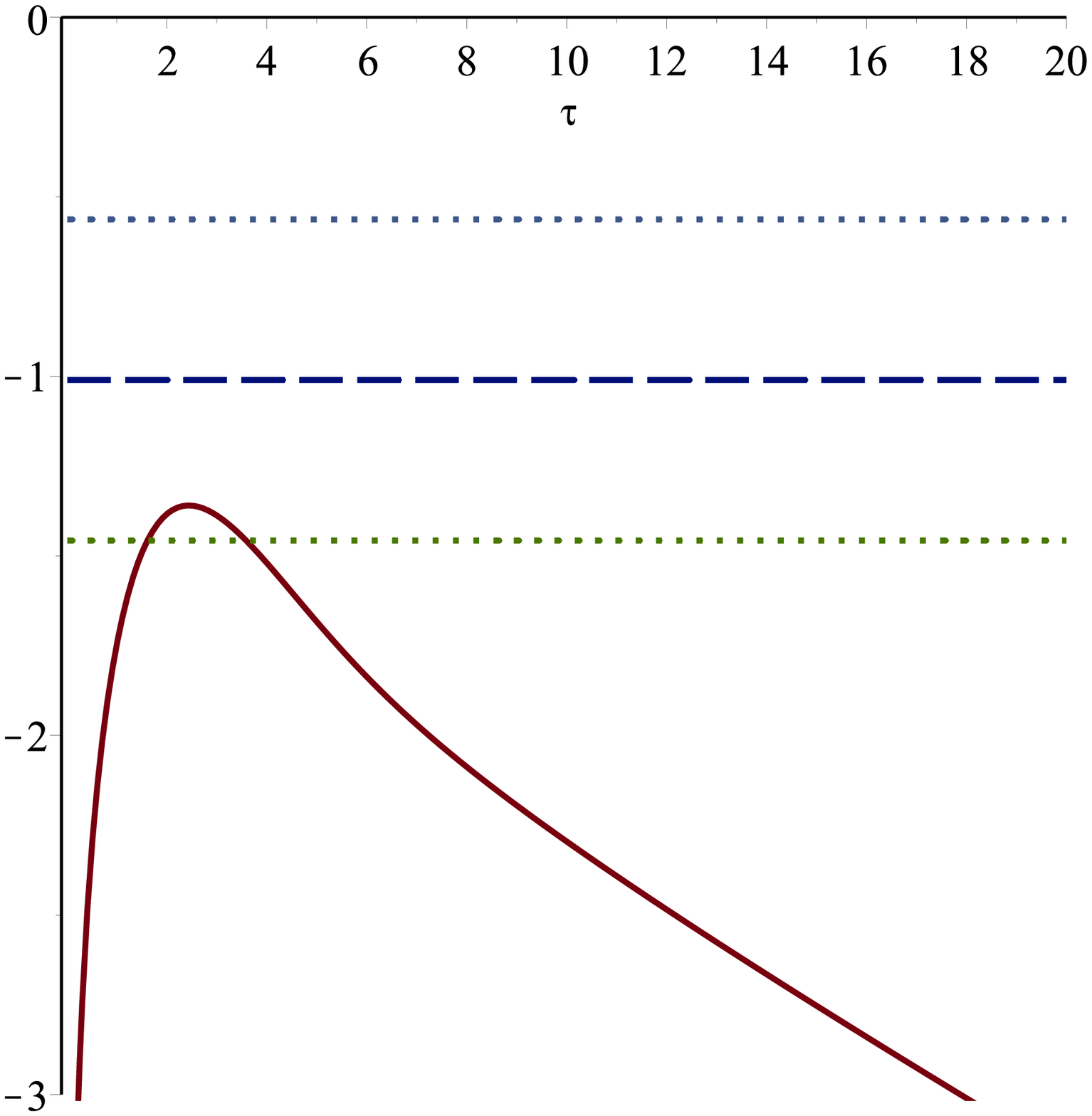, height=1.3in, width=1.3in}& \qquad\quad
\epsfig{file=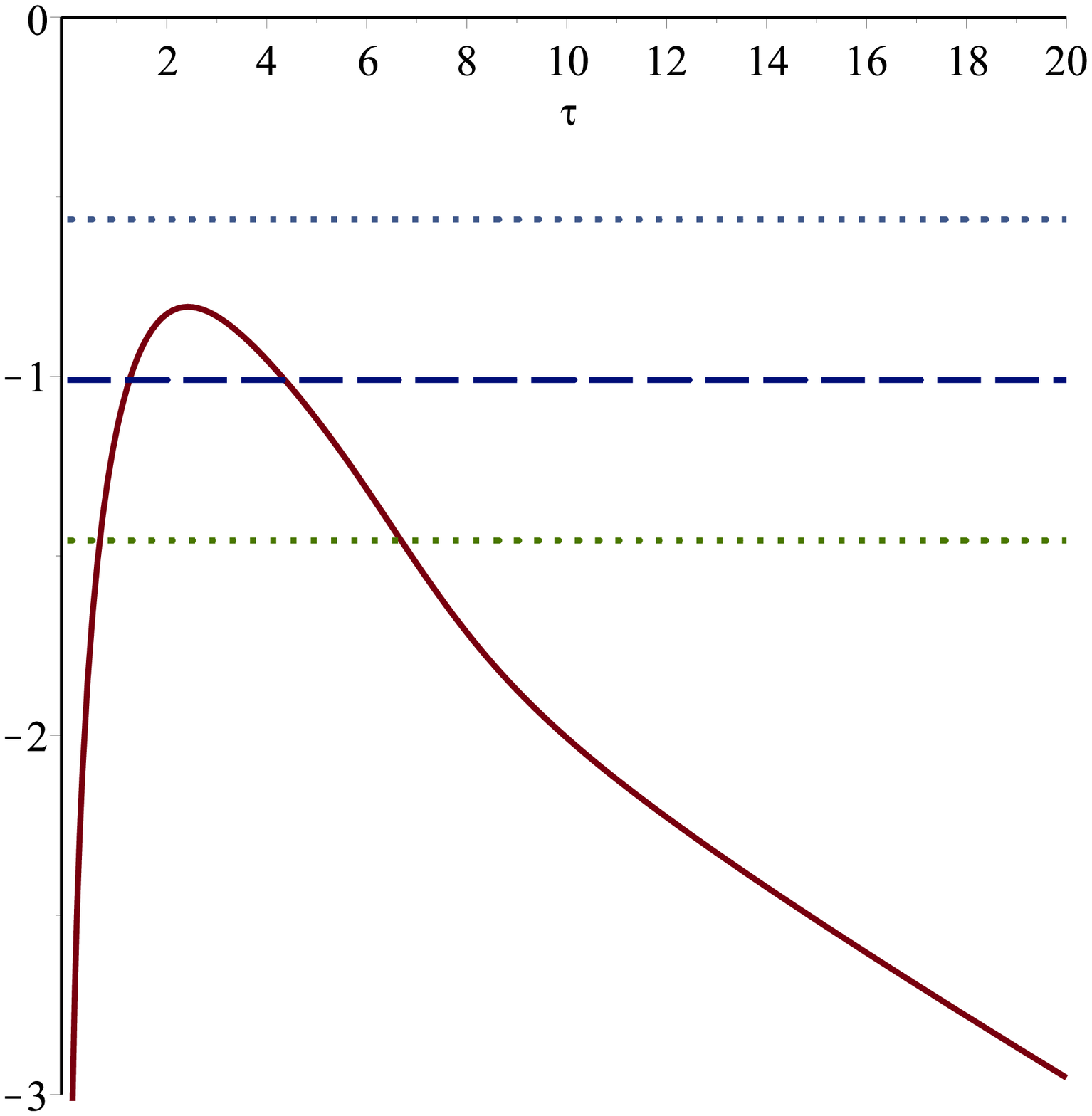, height=1.3in, width=1.3in} \\
\epsfig{file=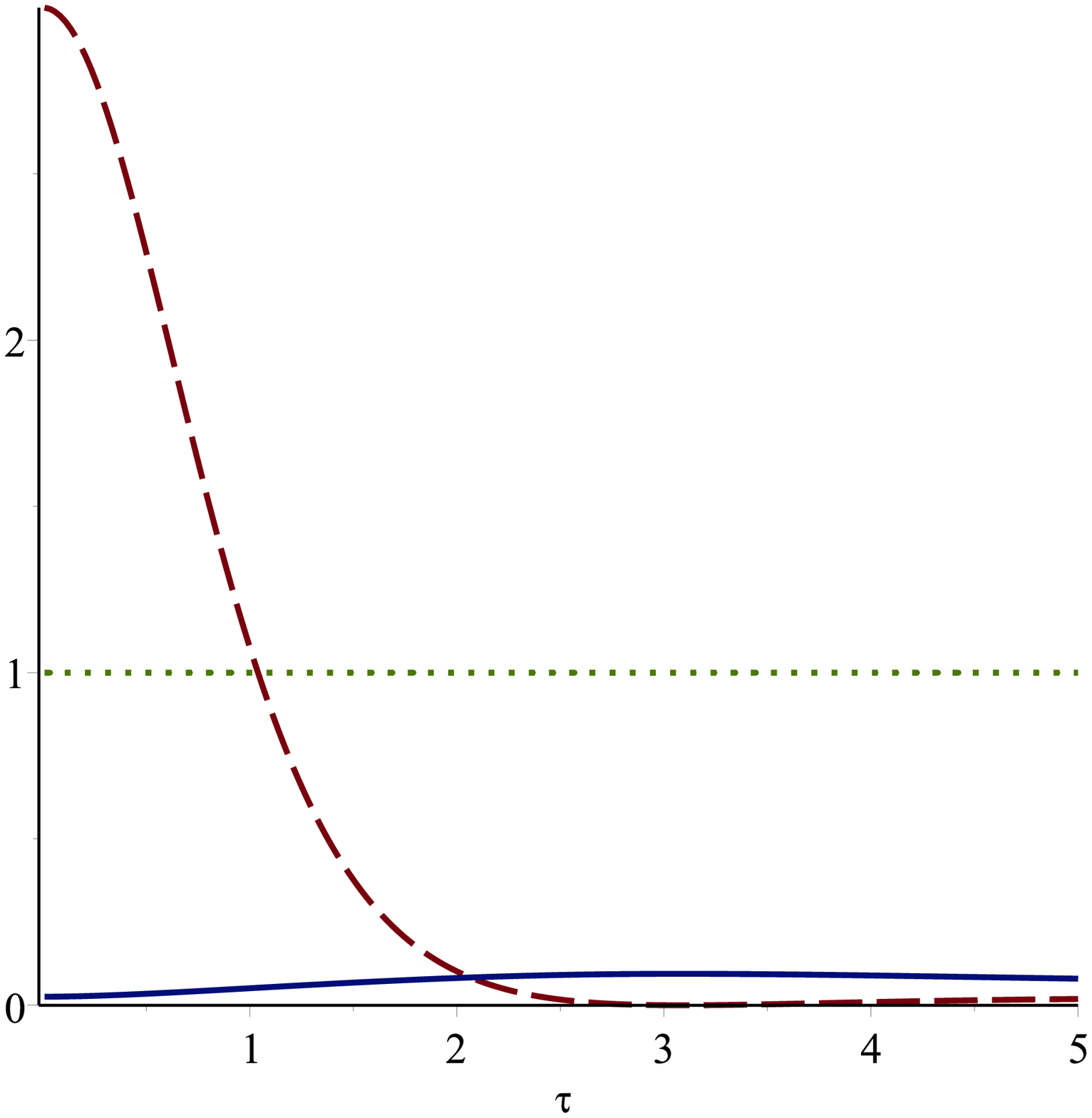, height=1.3in, width=1.3in}& \qquad\quad
\epsfig{file=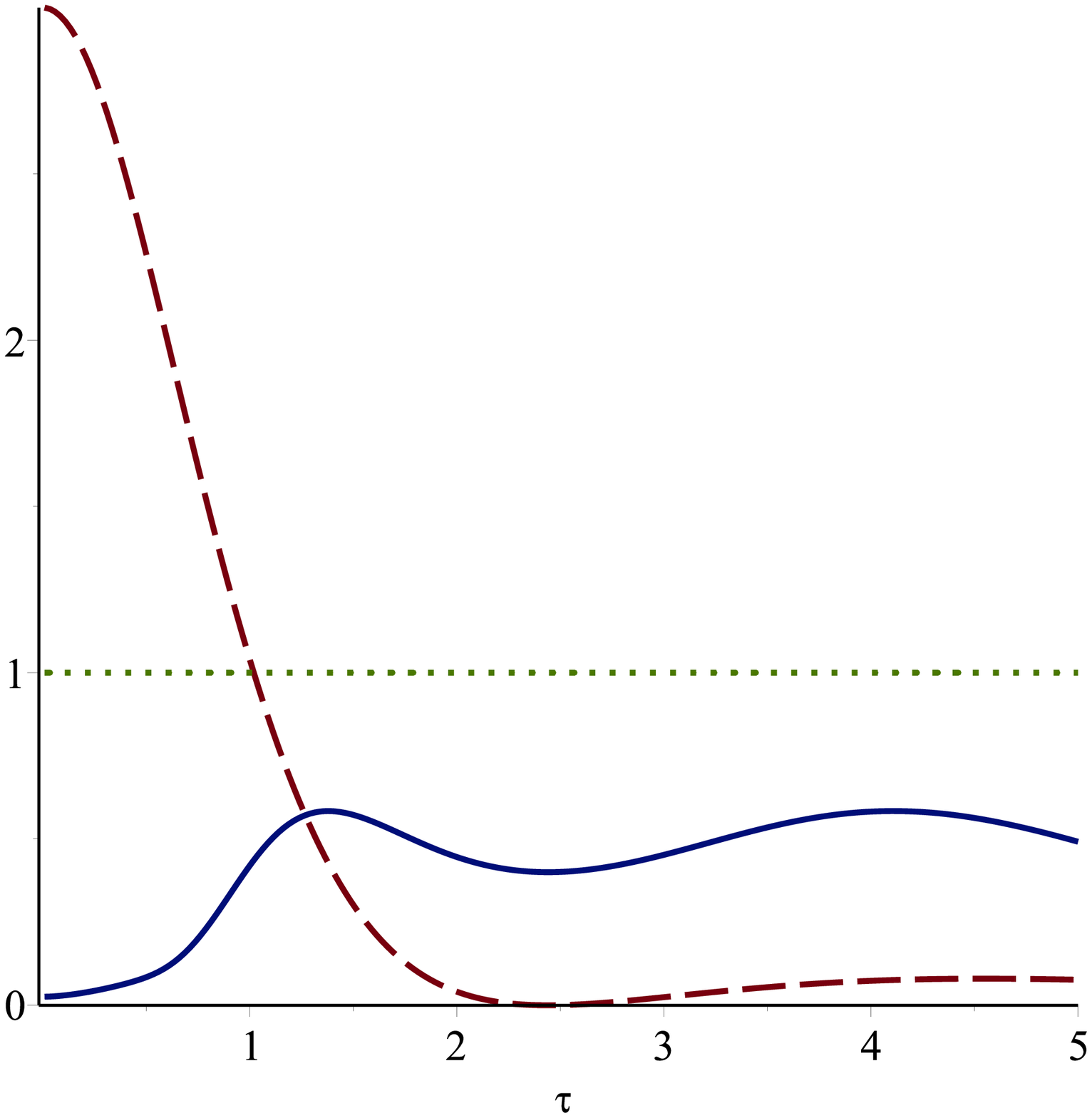, height=1.3in, width=1.3in}& \qquad\quad
\epsfig{file=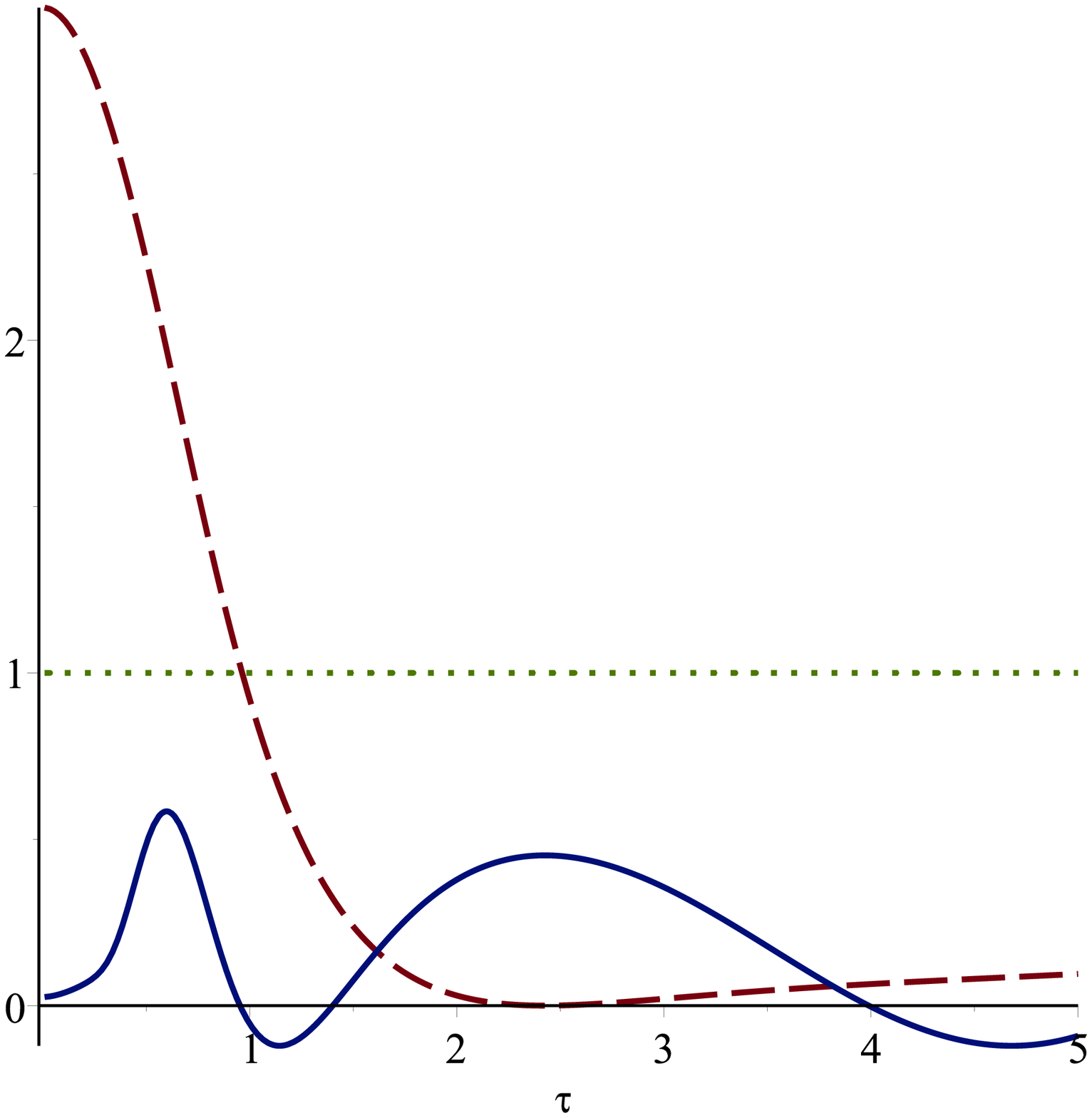, height=1.3in, width=1.3in}
\end{array}$
\end{center}
\caption{\small
The upper plots display the evolution of the scalar field $\varphi(\tau)$ for the three instructive choices $\varphi_0$: $\varphi_0=-2$ (left), $\varphi_0=-1$ (center) and $\varphi_0=-0.37$ (right). The dashed line is the central profile of the gaussian bump. The lower plots display the corresponding evolutions of the slow--roll parameters $\epsilon(\tau)$ (dashed) and $\eta(\tau)$ (continuous).}
\label{fig:eps_eta}
\end{figure}

Interestingly, the preferred choices of $\varphi_0$ that we identified working with WMAP9 and PLANCK 2013 data are consistent. As we have stressed, the choice $\gamma = 0.08$ was motivated by the naive correspondence between the mild exponential and an eventual spectral index $n_s \simeq 0.96$. For any choice of $\varphi_0$ we determined $M$ and $\delta$ optimizing the comparison with data. In detail:
\begin{itemize}
\item we minimized eq.~\eqref{chi_squared} analytically
 with respect to the normalization factor ${\cal M}$;
\item we then identified optimal choices for the parameter $\delta$ in eq.~\eqref{bessel} scanning the range $[-1.3, 1.3]$. The end result is a clear tendency to lock the first pre--inflationary peak to the feature present around $\ell=5$ in both WMAP9 and PLANCK 2013 data;
\end{itemize}

The reader will probably notice an amusing analogy between this type of analysis and Mean Field Theory. A nice by--product, which appears stable and independent of detailed tuning within the region of interest, is the emergence of a wide trough around $\ell=20$. However, the two--exponential spectra that we could generate from the models tend to approach rather slowly the attractor profile. Some ongoing likelihood tests \cite{gkmns} will say more on their actual significance.

\begin{figure}[ht]
\begin{center}$
\begin{array}{ccc}
\epsfig{file=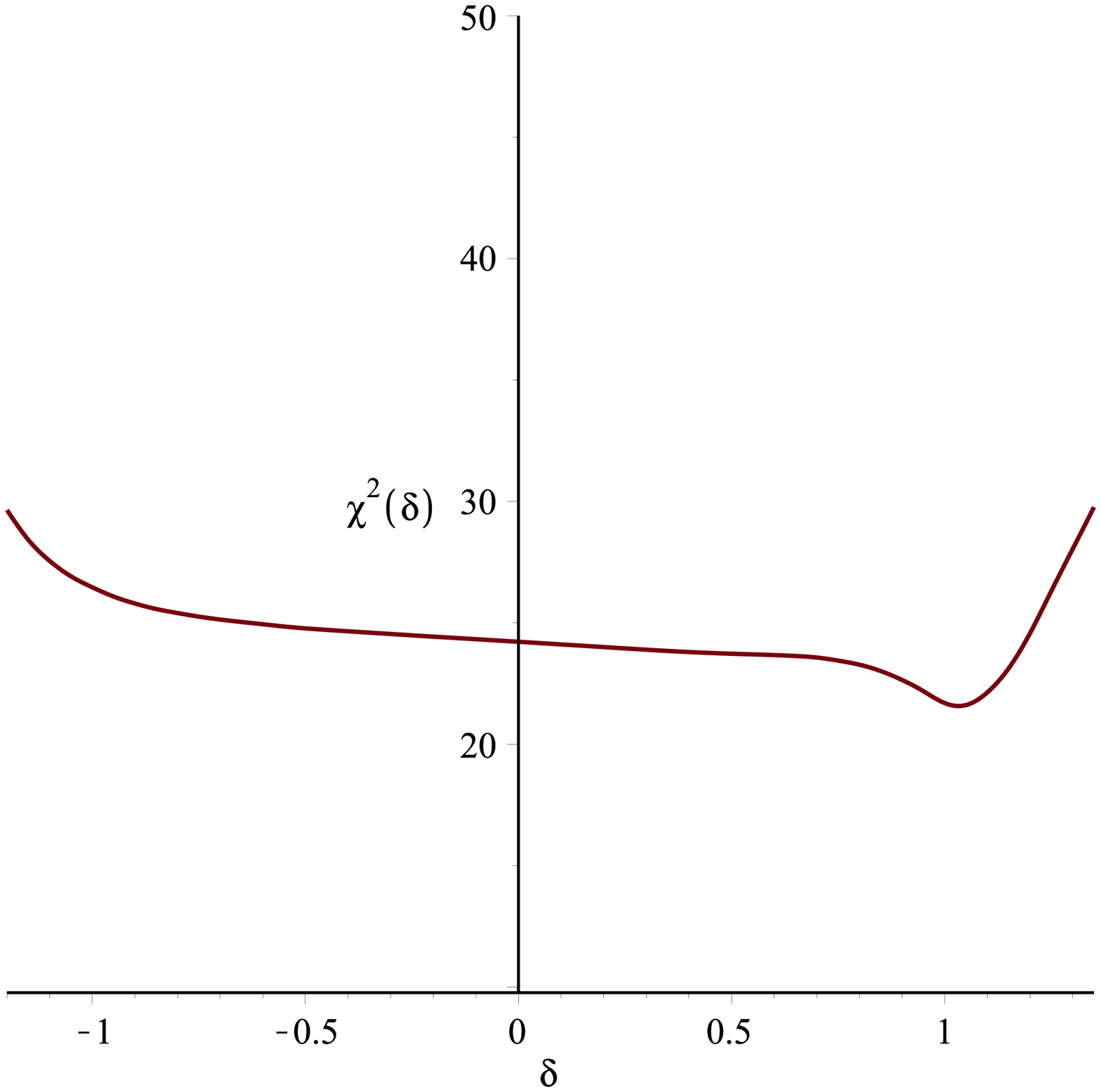, height=1.4in, width=1.4in}& \qquad\quad
\epsfig{file=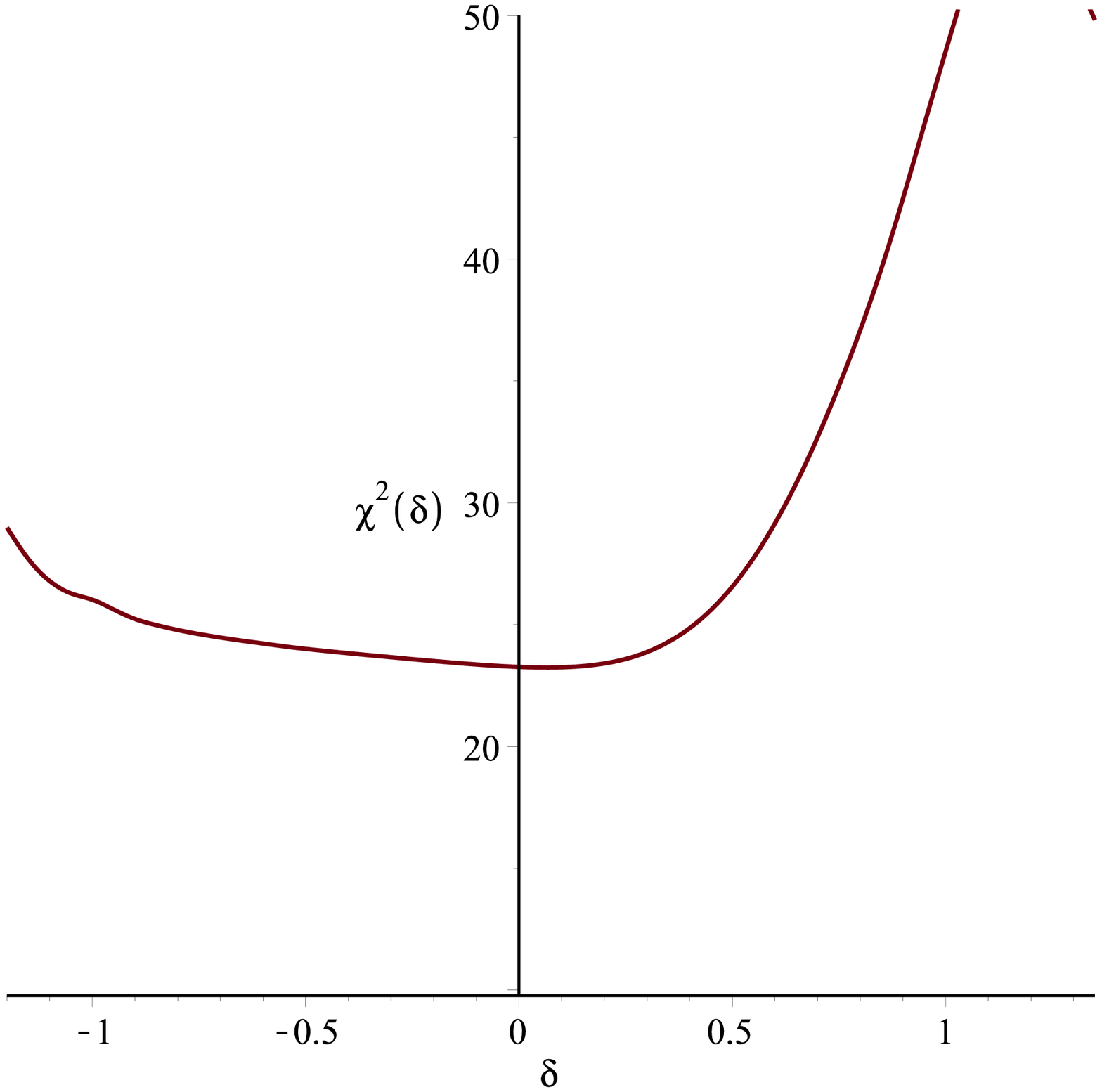, height=1.4in, width=1.4in}& \qquad\quad
\epsfig{file=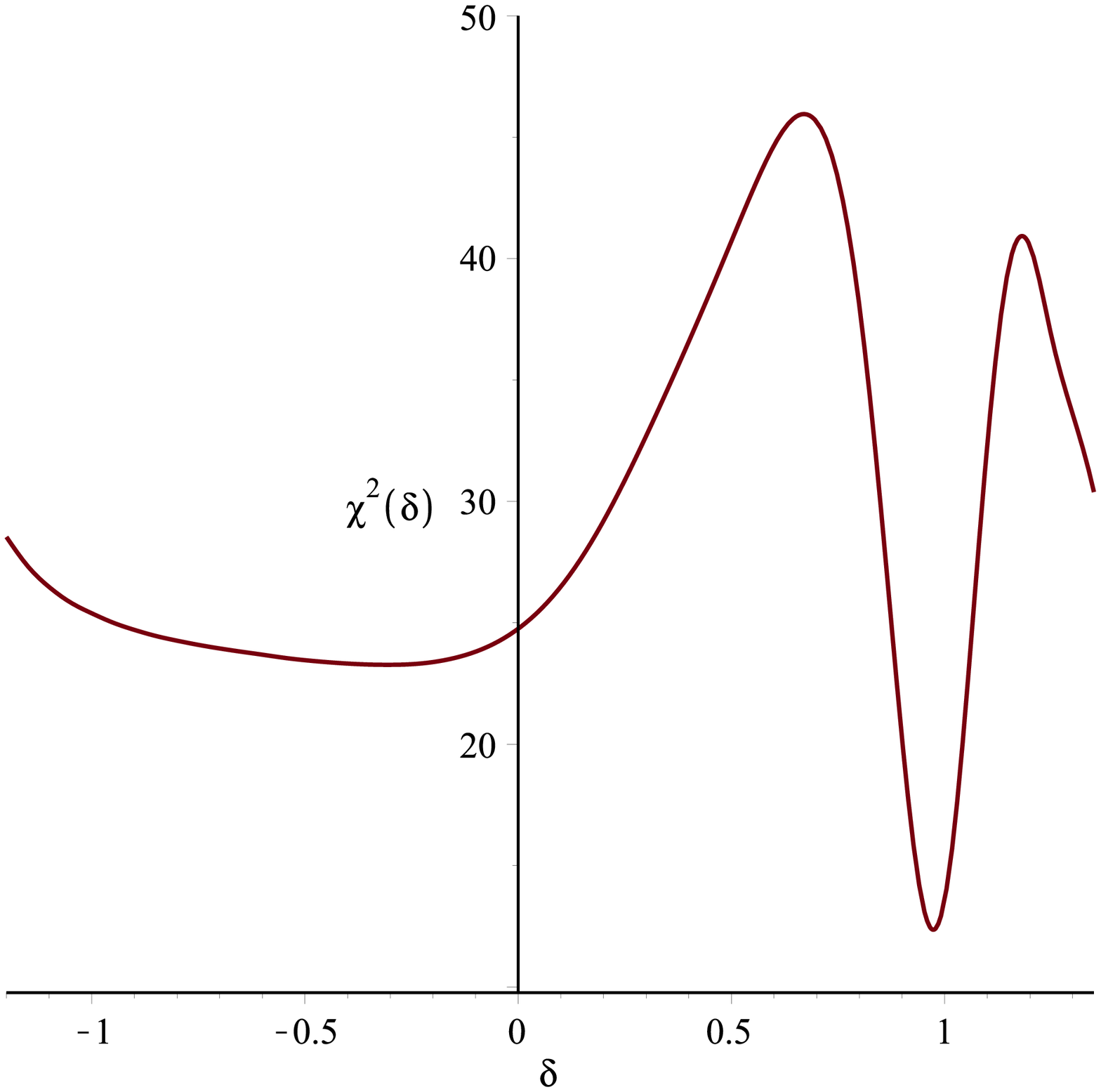, height=1.4in, width=1.4in}
\end{array}$
\end{center}
\caption{\small
The dependence of $\chi^{\,2}$, minimized with respect to $M$, on $\delta$, for three significant cases in fig.~\ref{fig:powerspectra}, corresponding to $\varphi_0=-2$ (left), $\varphi_0=-1$ (center), $\varphi_0=-0.37$ (right). The minima present in the first and last cases correlate the first pre--inflationary peak to the feature present in fig.~\ref{fig:raw_data} around $\ell=5$.}
\label{fig:chi2delta}
\end{figure}
\begin{figure}[ht]
\begin{center}$
\begin{array}{ccc}
\epsfig{file=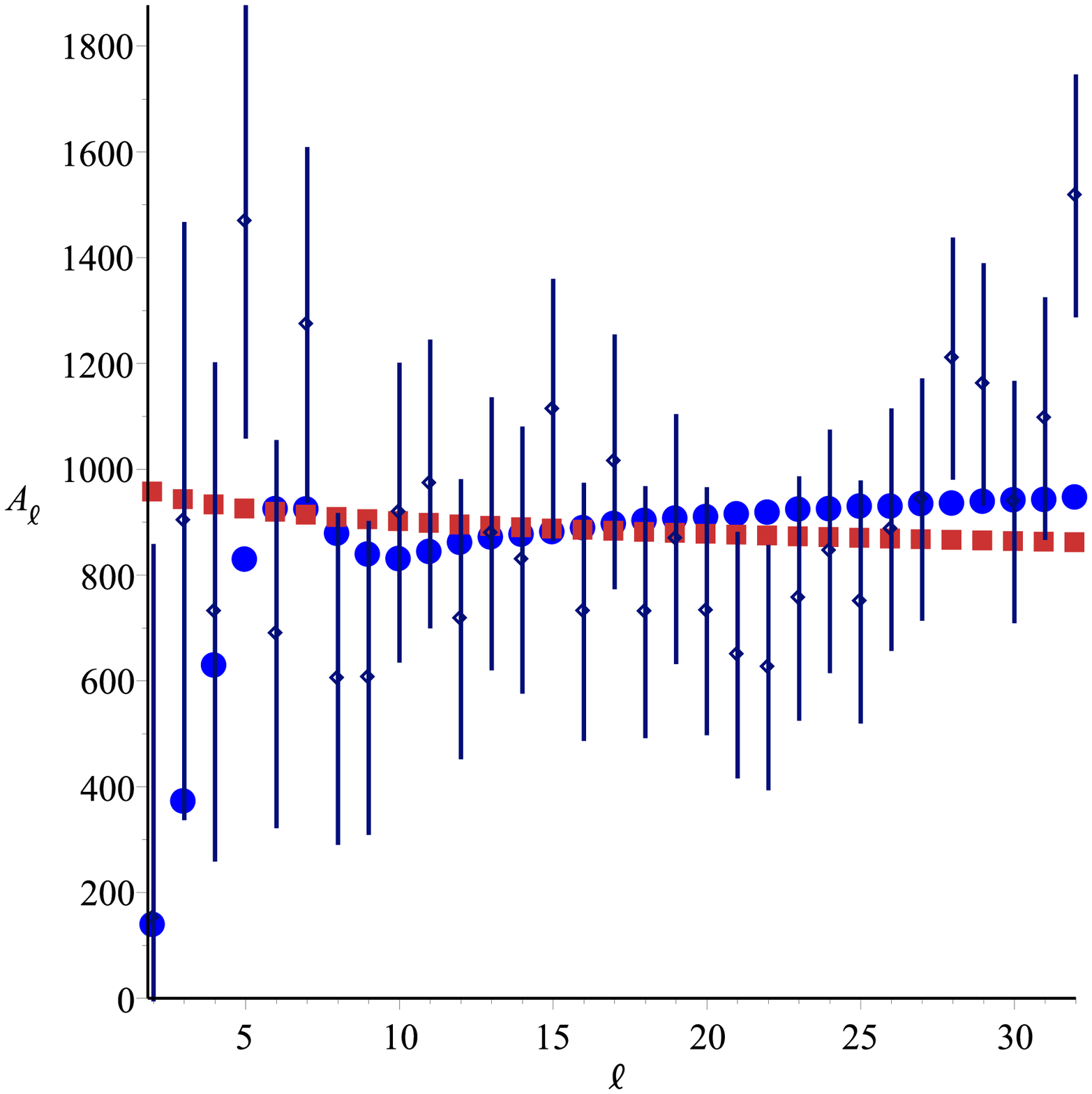, height=1.4in, width=1.4in}& \qquad\quad
\epsfig{file=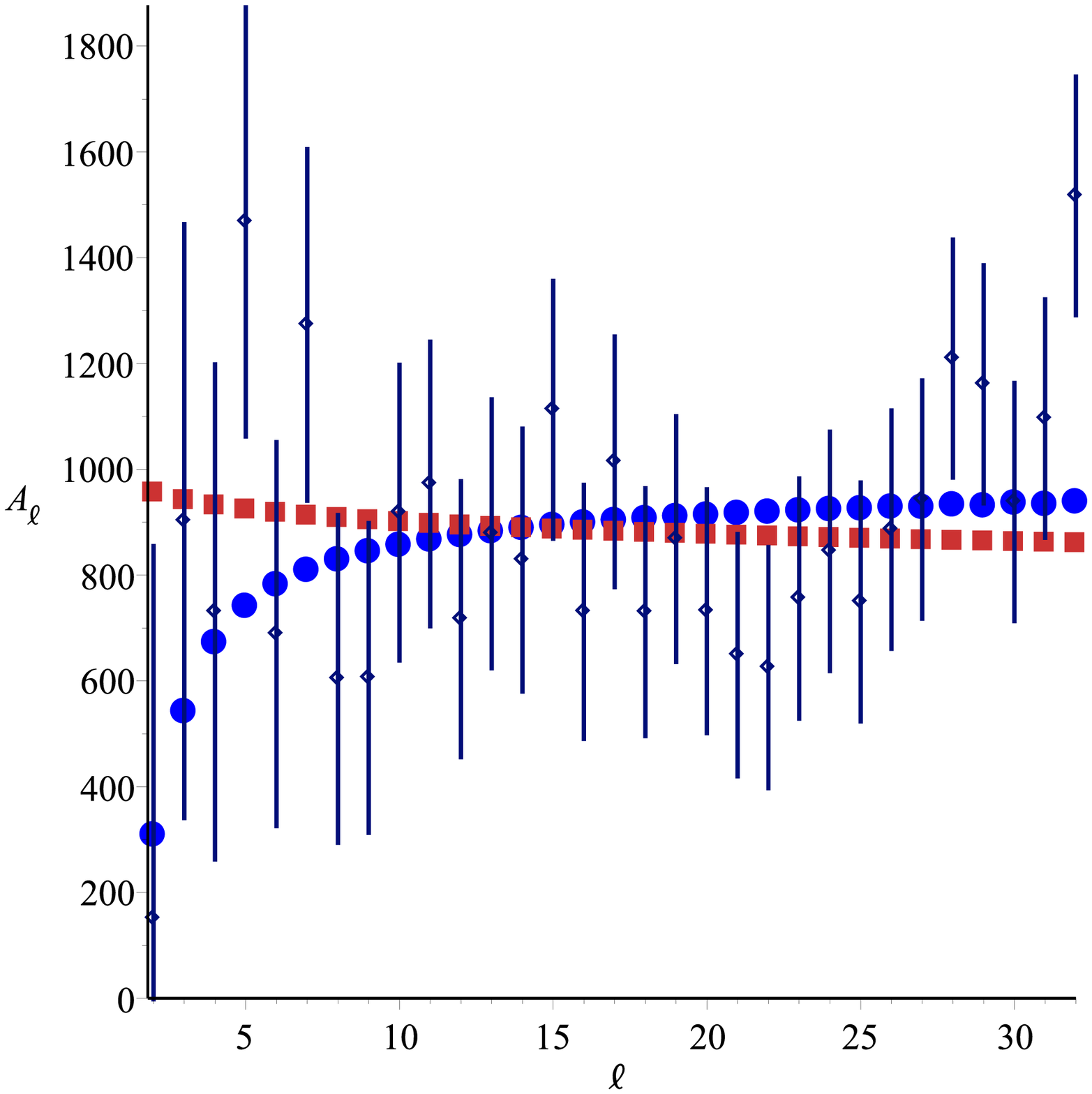, height=1.4in, width=1.4in}& \qquad\quad
\epsfig{file=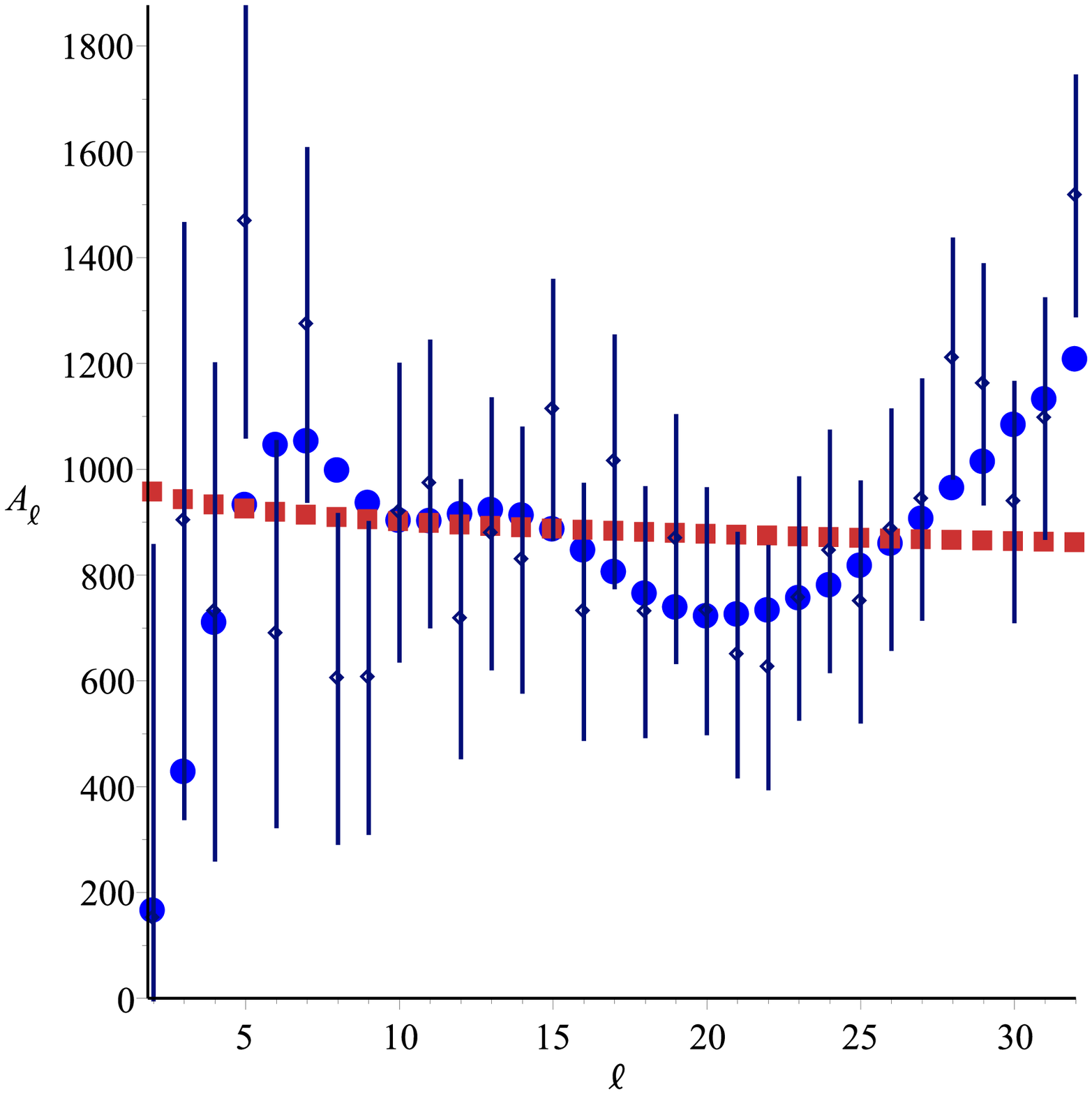, height=1.4in, width=1.4in}
\end{array}$
\end{center}
\caption{\small
Comparisons between the WMAP9 data, the $\Lambda$CDM angular power spectrum in the region $2 \leq \ell \leq 32$ (red) and the angular power spectra corresponding to $\varphi_0=-2$ (left, blue), $\varphi_0=-1$ (center, blue) and $\varphi_0=-0.37$ (right, blue).}
\label{fig:angular}
\end{figure}
Our first group of plots, collected in fig.~\ref{fig:powerspectra}, displays the evolution of the power spectrum as $\varphi_0$ is increased~\footnote{From $-4$ to $0$, in our conventions, which depend however on our choice of parametrizing initial conditions via exact solutions for the mild exponential. This contribution clearly dominates during the climbing phase and altogether for large negative values of $\varphi$, when the scalar reverts its motion far from the hard exponential. Increasing $\varphi_0$ within this range brings the turning point close to the bump, beyond it and then close to the exponential wall.}, making the scalar first feel the gaussian bump more closely, then overcome it and eventually start feeling more intensely the exponential wall. The examples collected in fig.~\ref{fig:eps_eta} reveal clearly the following pattern:
\begin{figure}[ht]
\begin{center}$
\begin{array}{ccc}
\epsfig{file=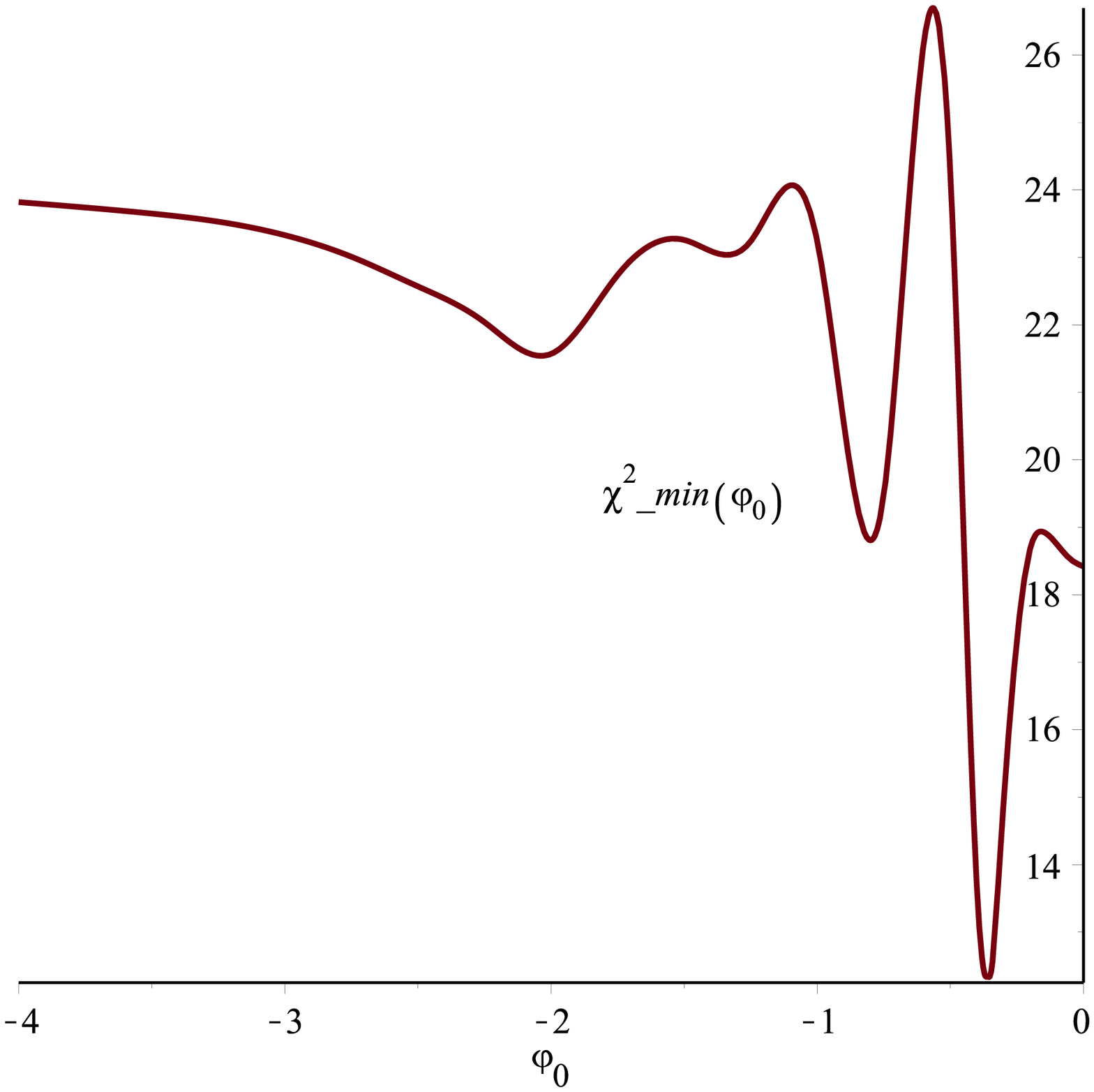, height=1.4in, width=1.4in}& \quad\qquad
\epsfig{file=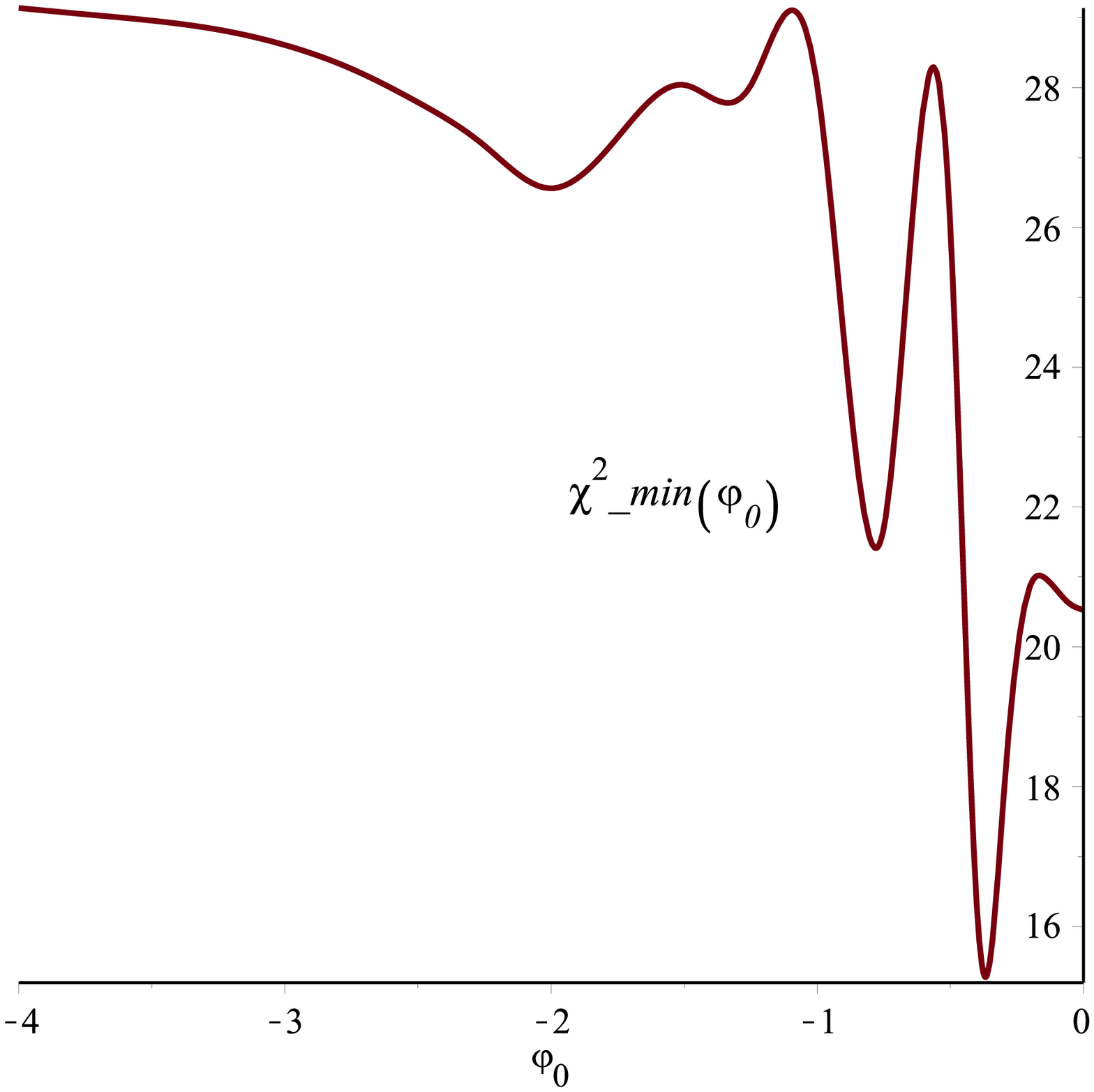, height=1.4in, width=1.4in}& \quad\qquad
\epsfig{file=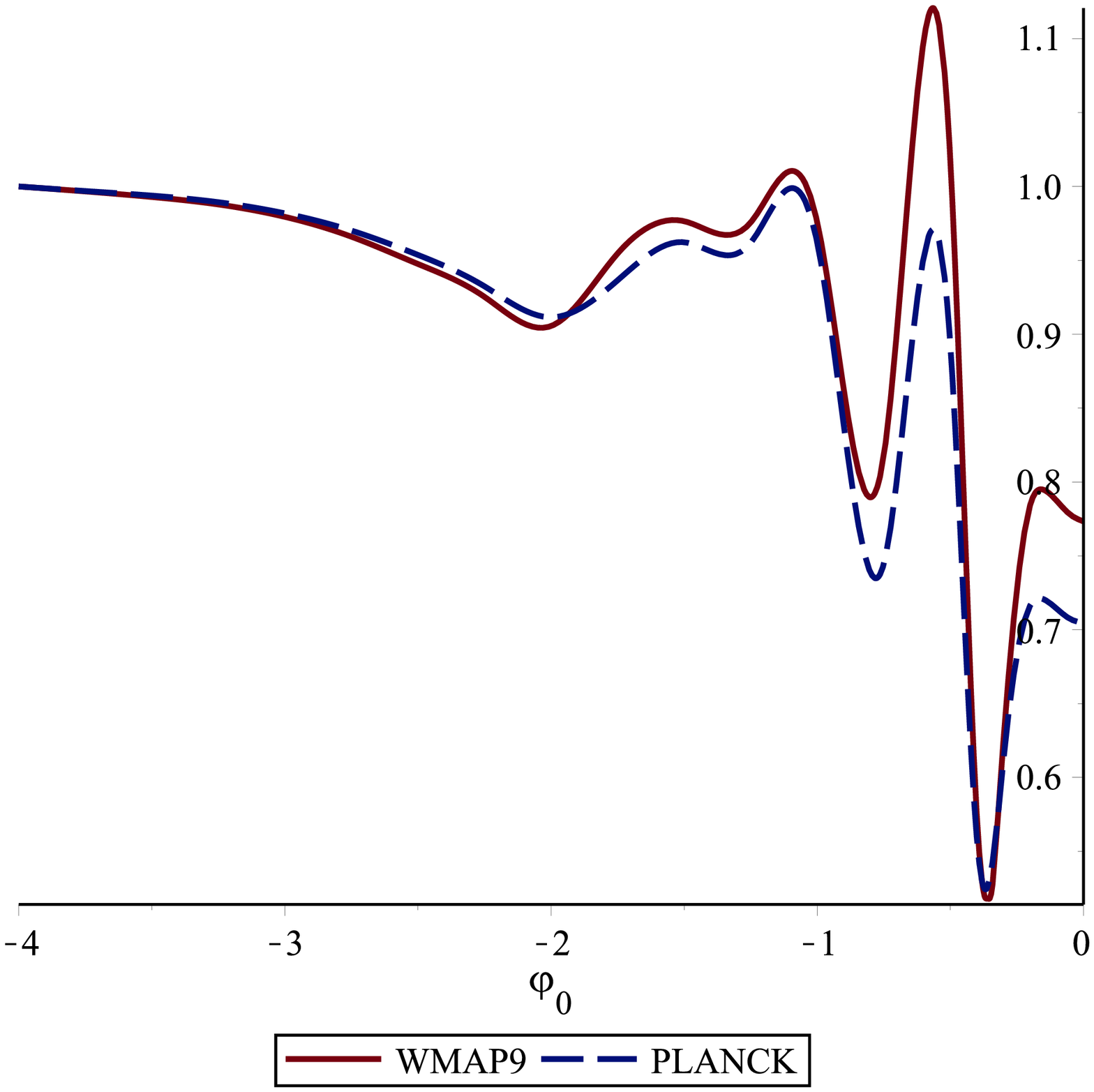, height=1.4in, width=1.4in}
\end{array}$
\end{center}
\caption{\small
The optimal values of $\chi^2$ obtained from WMAP9 data as a function of
 $\varphi_0$ (left), the corresponding optimal values obtained from PLANCK 2013 data (center) and the comparison between the two sets of data for WMAP9 (red) and PLANCK 2013 (blue, dashed), normalized with respect to their different values for $\varphi_0=-4$, which are respectively 29.14 and 23.82.}
\label{fig:angular_fit_double_bump}
\end{figure}
\begin{itemize}
\item for $-4<\varphi_0<-3$ the scalar hardly feels the bump. It is almost in slow--roll when it reverts its motion before it, so that the power spectrum is as in \cite{destri1,destri2}. As $\varphi_0$ is increased ($-3<\varphi_0<-1.75$) the scalar begins to feel the bump, its reversal occurs in a region where the curvature $V^{\,\prime\prime}$ of the potential is larger, and the pre--inflationary peak moves away from the slow--roll region, as in \cite{ks1}.
\item With a further increase of $\varphi_0$ ($-2<\varphi_0<-1$), the turning point moves closer to the bump, which makes the pre--inflationary peak smaller before it essentially disappears altogether. The end result is similar to what occurred close to the exponential wall in \cite{dkps} and in \cite{ks1,ks2}. However, the pre--inflationary peak builds up again, larger and larger, as the scalar begins to linger near the top of the bump, thus exploring a region where $V^{\,\prime\prime}$ is smaller. This type of effect was expected, and motivated us initially to introduce a gaussian bump.
\item With a further increase of $\varphi_0$ ($\varphi_0>-1$), the scalar climbs up beyond the bump, and now an interesting new pattern emerges. A first pre--inflationary peak builds up as the scalar overcomes the gaussian bump, around which $V^{\,\prime\prime}$ is small and negative. It is followed by a mild trough, consistent with the fact that the turning point is beyond the bump, in a region where $V^{\,\prime\prime}$ is relatively large and positive, so that the corresponding perturbations do not grow efficiently, by a second peak that builds up as the scalar overcomes the bump again during the ensuing descent and by a deeper trough as it speeds up after it. The spectrum then approaches slowly the attractor shape of \cite{cm} as slow--roll is finally attained.
\end{itemize}

Our next group of plots, in fig.~\ref{fig:chi2delta}, displays the dependence of $\chi^2$ on $\delta$ for three significant cases of fig.~\ref{fig:powerspectra}. The first corresponds to $\varphi_0=-2$, where a single peak is present well apart from the attractor profile. The second corresponds to $\varphi_0=-1$, where the peak has disappeared, and finally the third corresponds to $\varphi_0=-0.37$, which lies well within the double--peak region. Notice the presence of narrow minima, shallow in the first case and rather deep in the third, which correlate the pre--inflationary peaks to the feature that the data display around $\ell=5$. On the other hand, the wide minimum that is present in the second case correlates the low quadrupole to the growing portion of the power spectrum.

Our next group of plots, in fig.~\ref{fig:angular}, displays the angular power spectra for the three choices of $\varphi_0$ corresponding to fig.~\ref{fig:chi2delta}.
\begin{figure}[ht]
\begin{center}$
\begin{array}{ccc}
\epsfig{file=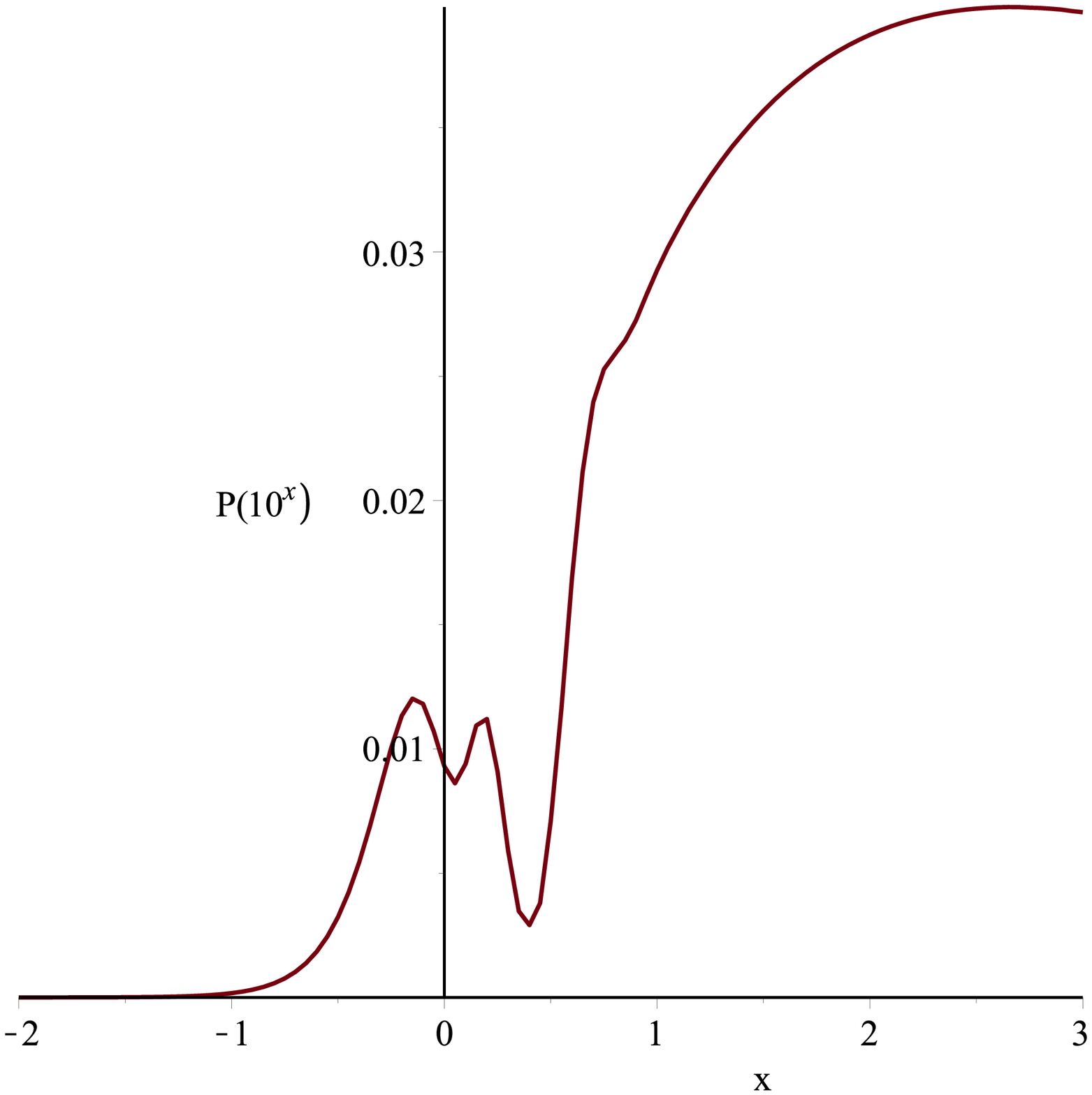, height=1.4in, width=1.4in}& \quad\qquad
\epsfig{file=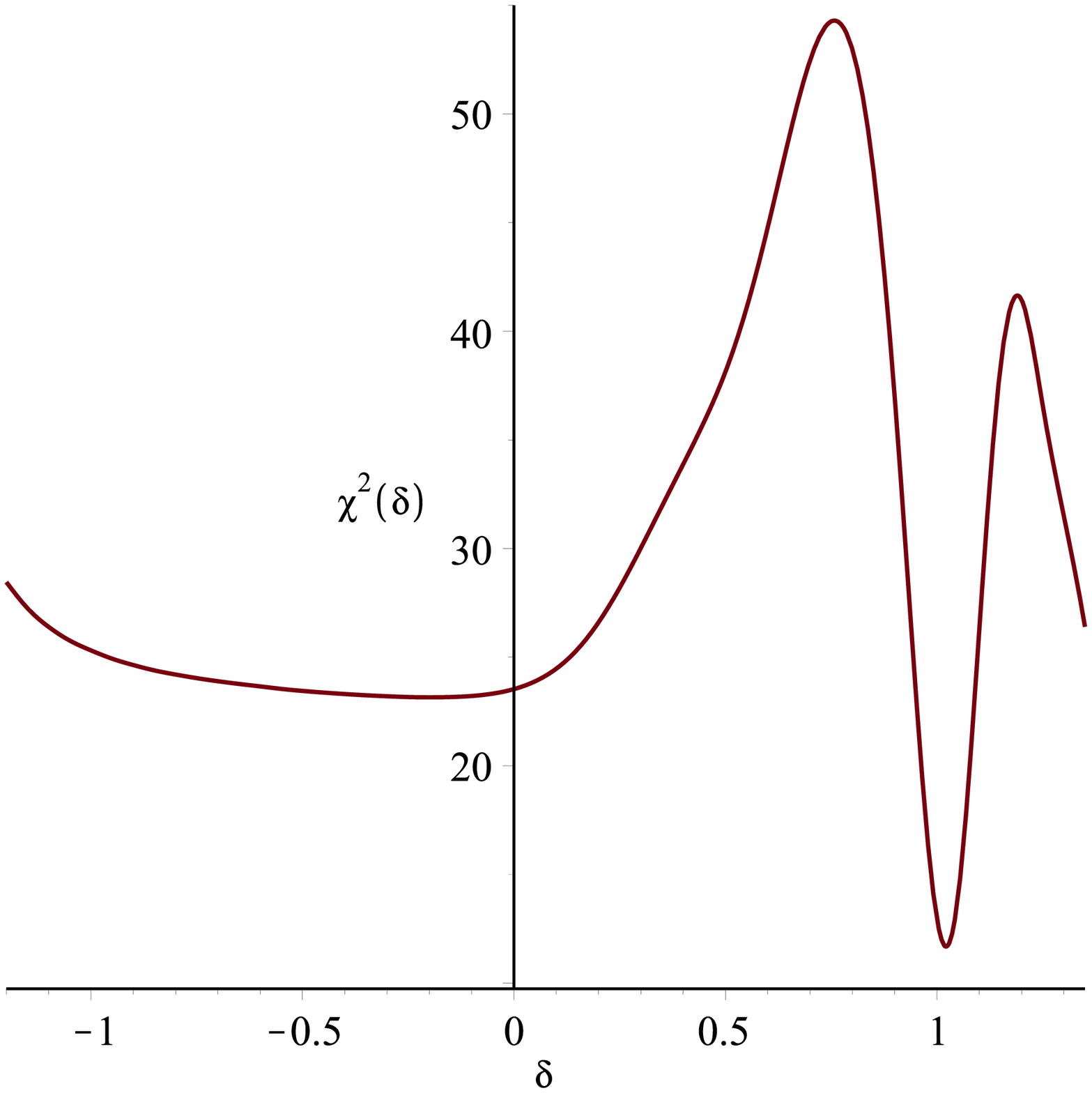, height=1.4in, width=1.4in}& \quad\qquad
\epsfig{file=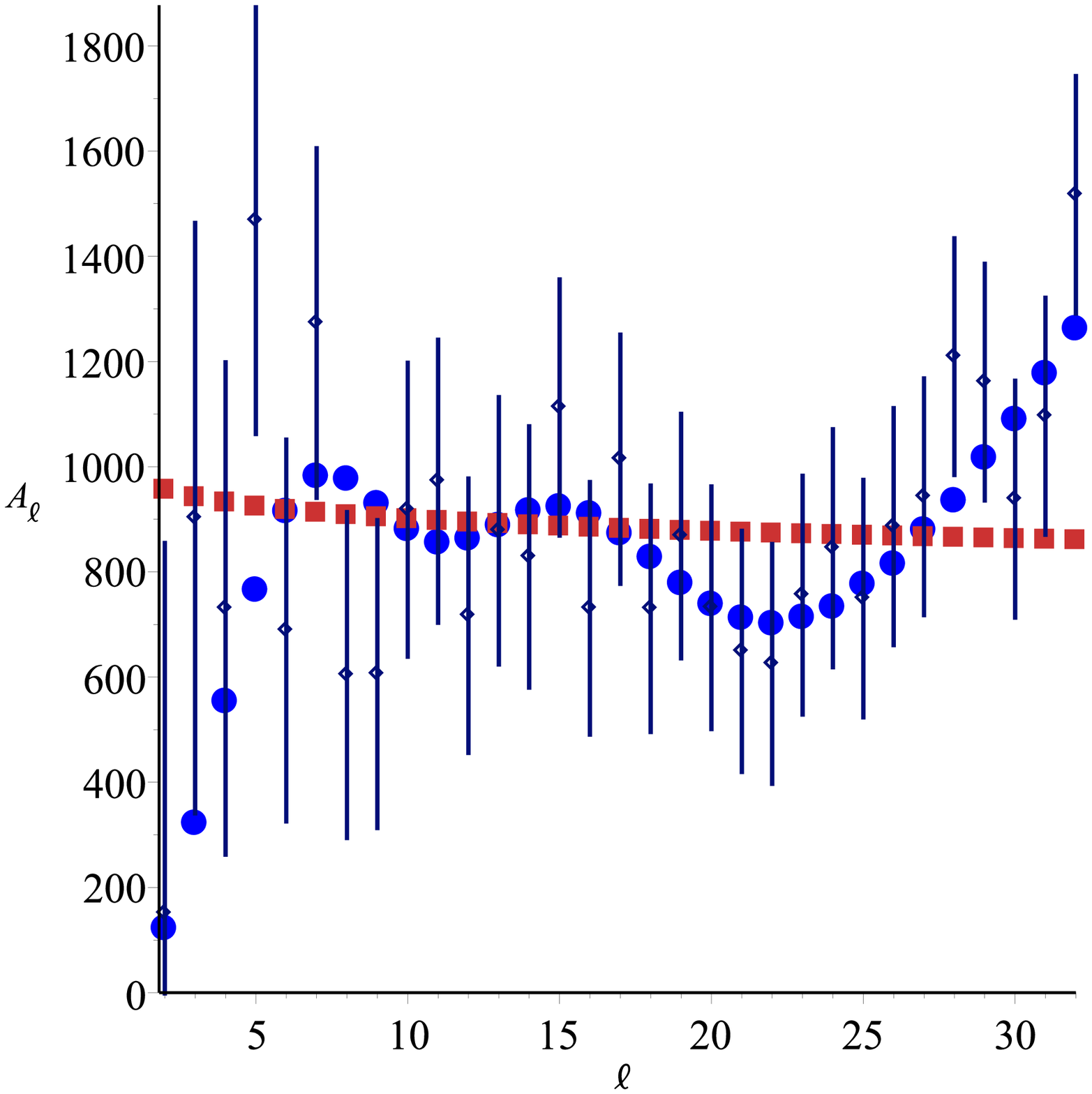, height=1.4in, width=1.4in}
\end{array}$
\end{center}
\caption{\small
The power spectrum of scalar perturbations (left, up to an overall normalization) for the two--exponential system with $\gamma=0.08$, a gaussian bump with $(a_1,a_2,a_3)=(0.086,6,0.9)$ and $\varphi_0=-0.2$. The other figures display the corresponding $\chi^{\,2}(\delta)$ (center) and the angular power spectrum compared with WMAP9 data and with the $\Lambda$CDM result (right). With WMAP9 data this system led to $\chi^{\,2}=11.68$, the lowest value that we could obtain, about $46\%$ of 25.5, the $\Lambda$CDM value. With PLANCK 2013 data the result was $\chi^{\,2}=14$, again about $46\%$ of 30.3, the corresponding $\Lambda$CDM value. }
\label{fig:2expbest}
\end{figure}

Our next group of plots, in fig.~\ref{fig:angular_fit_double_bump}, displays the dependence on $\varphi_0$ of the minimum $\chi^2(\delta)$ obtained from the power spectra of fig.~\ref{fig:powerspectra} adjusting $M$ and $\delta$. The left curve refers to WMAP9 with standard galactic mask, while the central one refers to PLANCK 2013, again with standard galactic mask. Finally, the right plot compares the two figures, after adjusting their overall normalizations so that they coincide for $\varphi_0=-4$. Notice that the values of $\chi^2(\delta)$ are consistently higher for PLANCK 2013, but the optimal choices fall essentially on top of each other for the example chosen, and obtain in this example for $\varphi_0 \simeq -0.35$.

\vskip 12pt

\section{\sc  Discussion}\label{sec:conclusions}

The main thrust of this work was to take seriously some low--$\ell$ anomalies of the CMB angular power spectrum, under the spell of some clues provided by String Theory. Referring to fig.~\ref{fig:raw_data}, these anomalies include a sizable quadrupole reduction and some oscillations, with a first peak around $\ell=5$ and a wide depression around $\ell=20$.

A notable class of string vacua with ``brane supersymmetry breaking'', where supersymmetry is broken at the scale of string excitations, involves an exponential potential that is just too steep for a scalar field to emerge from the initial singularity while descending it. \emph{The overall lesson of the resulting picture is that an early climbing phase and a bounce could have preceded the onset of inflation.} This type of dynamics brings along, in the primordial power spectrum, an infrared depression and pre--inflationary peaks. A further step, which is not logically implied, was to look for relics of this transition in the CMB. This would require that the phenomena be accessible to us via the largest wavelengths that are reaching us at present, which would appear compatible with a short inflation with $N \sim 60\,e$--folds. An early phase of fast--roll would then translate almost verbatim into a low--$\ell$ depression of the angular power spectrum, and thus into a low quadrupole. Witnessing via the CMB some relics of the onset of inflation would be most exciting: here we explored further this option, taking the angular power spectrum as a guiding principle, although a short inflation creates some tension with the original motivations \cite{inflation1} - \cite{inflation13} for this early phase. There are also competing effects: for instance, in this type of picture the ripples of the CMB might bear signs of inhomogeneities, especially for the first few multipoles, but at the same time their effect would be counteracted by their small resolution. Other CMB anomalies that would seem to point in this direction were actually noticed and are under investigation, including a peculiar directional alignment between quadrupole and octupole contributions \cite{pgn}. Moreover, one would be confronted, intriguingly, with another sort of cosmic coincidence, in addition to the recent onset of a novel phase of accelerated expansion.

In the preceding section we have displayed some power spectra obtained from the double exponential model with a gaussian bump located in the neighborhood of the exponential wall, together with corresponding plots of $\chi^{\,2}(\delta)$ and of low--$\ell$ angular power spectra. Two--exponential models provide the simplest setting to combine a bounce and a subsequent approach to a slow--roll regime. They bring about a typical signature, a pre--inflationary peak in the power spectrum of scalar perturbations, which can enhance to a double peak in the presence of a gaussian bump. In fig.~\ref{fig:angular_fit_double_bump} we have displayed the dependence on $\varphi_0$, the parameter that sizes the intensity of the bounce, of the optimized $\chi^{\,2}$ obtained working with WMAP9 and PLANCK 2013 data. The results are nicely consistent, and display narrow minima in the region where a double pre--inflationary peak is present. These minima actually superpose once the overall normalizations are adjusted so that the curves coincide for $\varphi_0=-4$, where the scalar feels neither the bump nor the exponential wall.

We have explored extensively the parameter space of two--exponential systems giving rise to two nearby pre--inflationary peaks, which appear favored in the comparison with the data of fig.~\ref{fig:raw_data}. We cannot claim to have exhausted the interesting region, but for completeness we have collected in fig.~\ref{fig:2expbest} the power spectrum, $\chi^2(\delta)$ and the angular power spectrum for a choice of parameters giving rise to $\chi^{\,2}=11.68$, the lowest value that we could obtain so far.
\begin{figure}[ht]
\begin{center}$
\begin{array}{cccc}
\epsfig{file=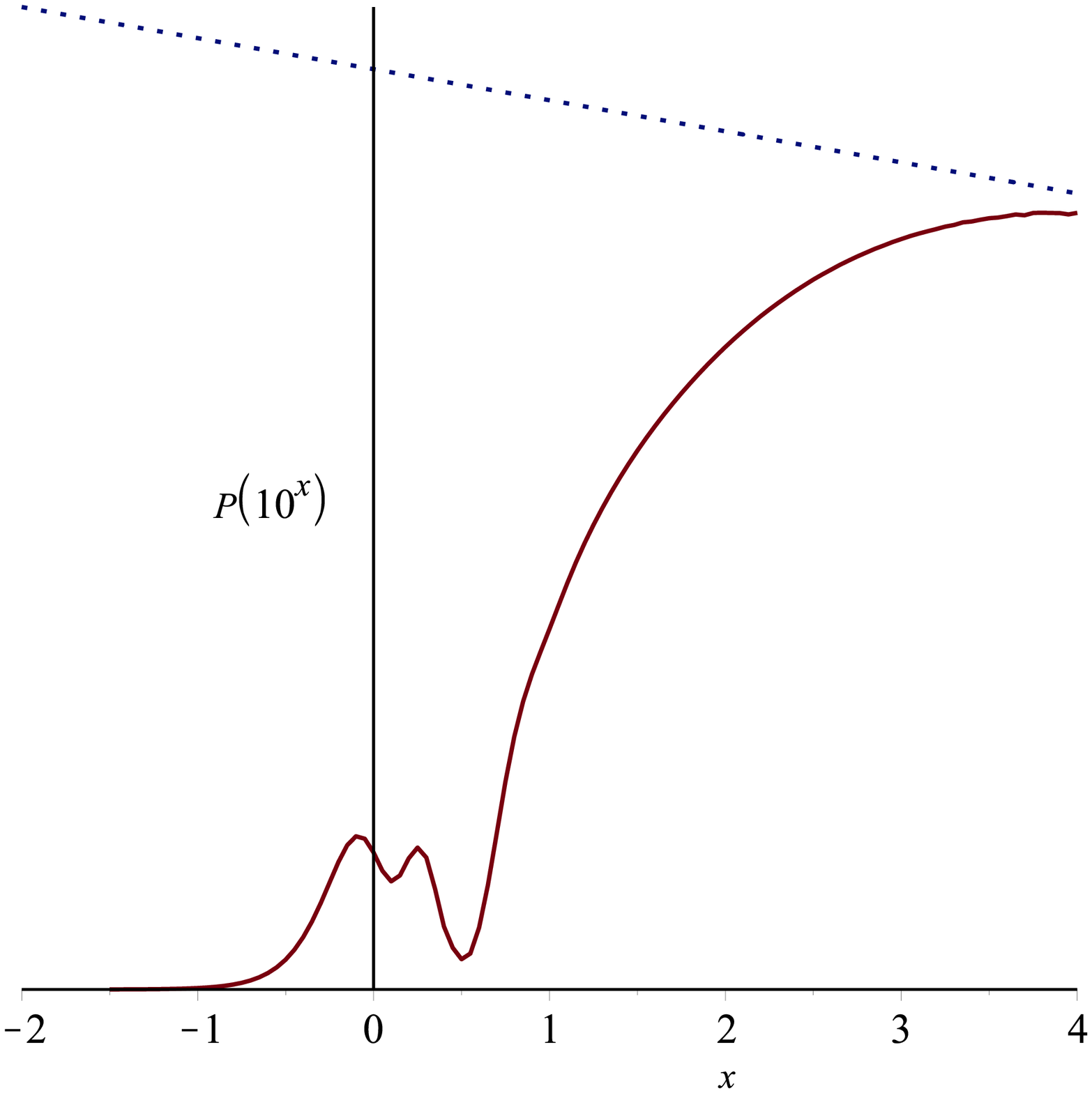, height=1.3in, width=1.3in}& \quad
\epsfig{file=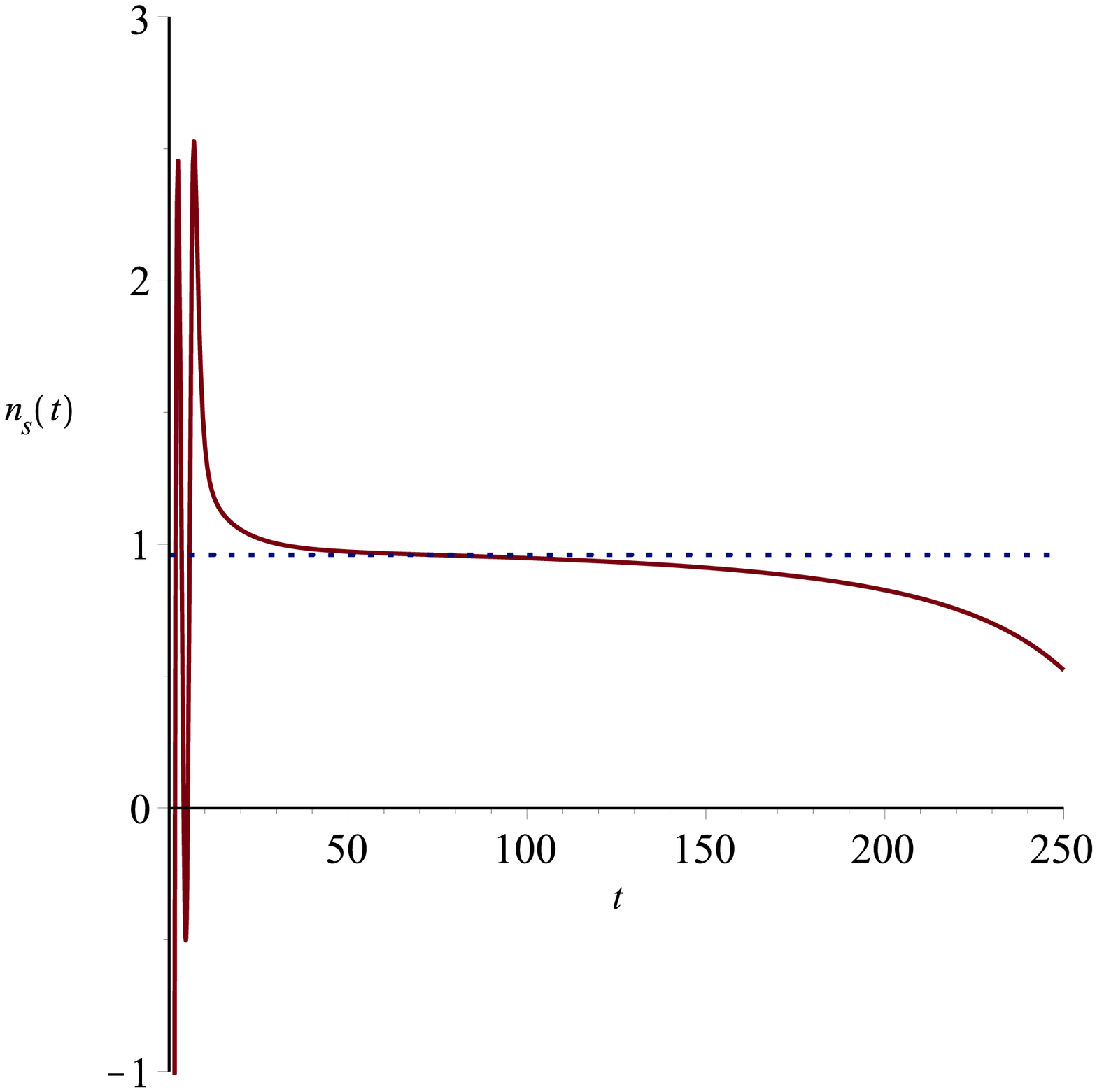, height=1.3in, width=1.3in}& \quad
\epsfig{file=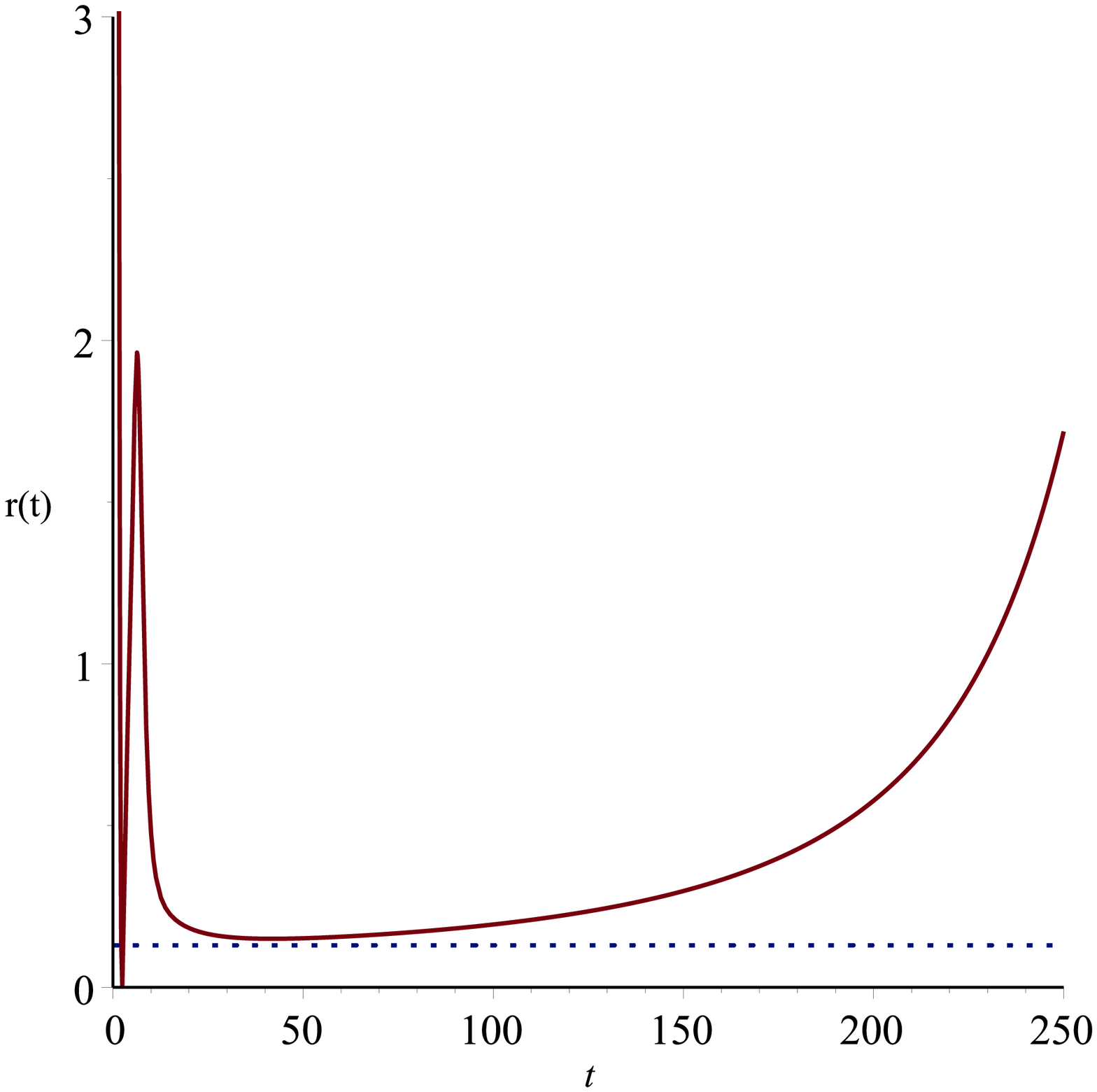, height=1.3in, width=1.3in}& \quad
\epsfig{file=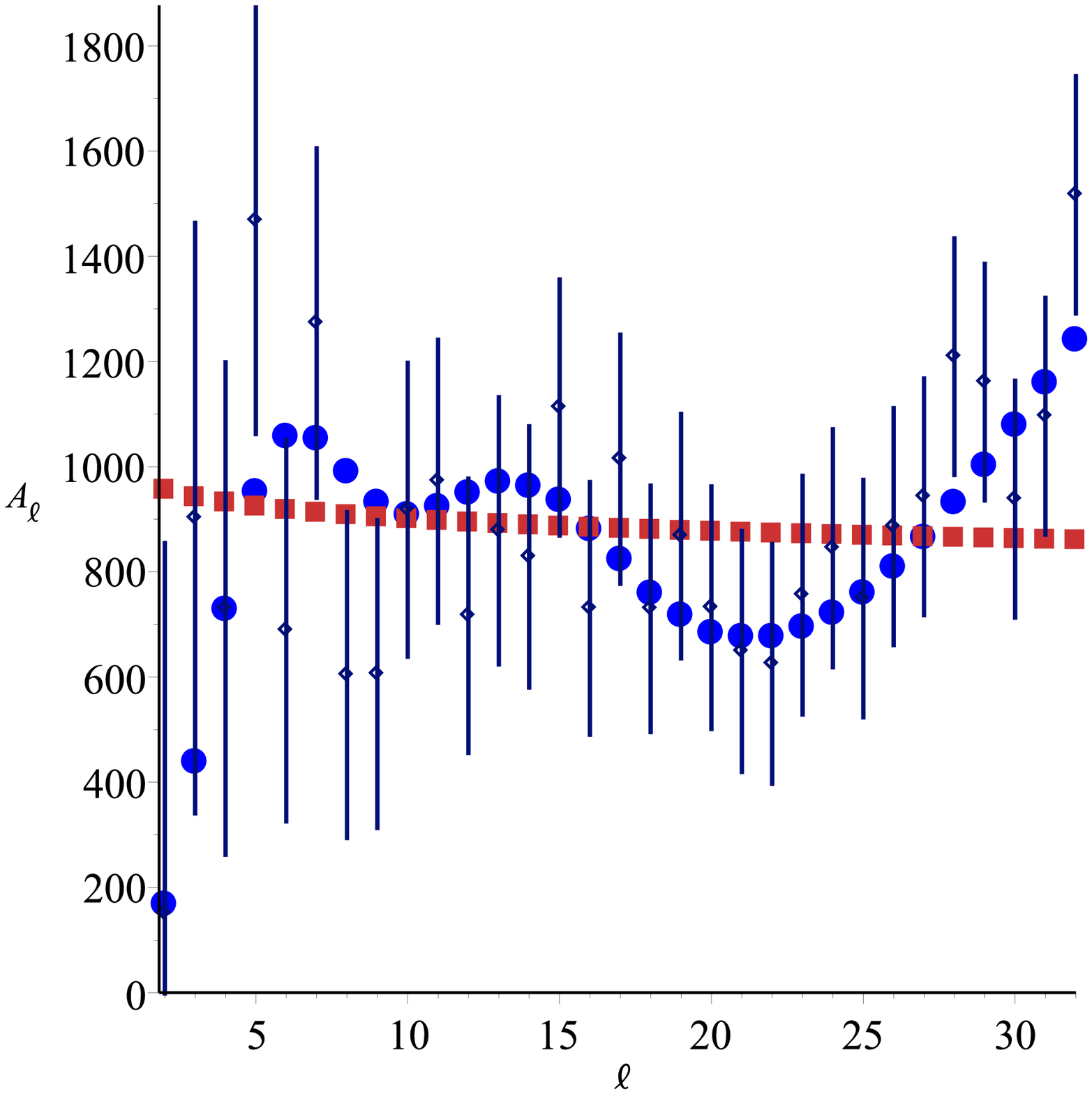, height=1.3in, width=1.3in}
\end{array}$
\end{center}
\caption{\small
Power spectrum of scalar perturbations, $n_s(t)$ and $r(t)$ during inflation, and the low--$\ell$ angular power spectrum for a Starobinsky--like system combined with the potential of eq.~\eqref{2expgauss} (with mild exponential term depressed by a factor 2). For this model $\gamma=0.08$, $(a_1,a_2,a_3)=(0.09,6,0.9)$, $\Delta=18$. The least $\chi^2(\delta)$ is 12.49, and there are 60 $e$--folds of inflation.}
\label{fig:starobinsky}
\end{figure}

We have also explored, although to a lesser extent, combinations of the potentials of eq.~\eqref{2expgauss} and Starobinsky--like terms
\beq
V \ = \ V_0 \ \left[1 - e^{\,-\, \alpha \, \left(\varphi \,+\, \Delta\right)} \right]^2 \ ,
\eeq{starobinsky}
where $\alpha =2/3$ in this notation. Our motivation was to try to combine an early climbing phase with an eventual graceful exit, while keeping $n_s$ around 0.96 and the tensor--to--scalar ratio $r$ around the current limit of about 0.13 \cite{planckbicep}. The computations are harder and had to be performed in cosmic time, but the resulting power spectra are qualitatively similar to those of two--exponential systems, as pertains to local effects that originate in the vicinity of the exponential wall. Double peaks emerge again, and adjusting $\Delta$ one can combine them nicely with 60 $e$--folds or so of inflation. Fig.~\ref{fig:starobinsky} displays the power spectrum, the behavior of $n_s$ and $r$ and the angular power spectrum of one of these systems obtained depressing the mild exponential term $\exp(2\gamma\varphi)$ in eq.~\eqref{2expgauss} by a factor of two in order to reduce $r(t)$. The model leads to 60 $e$--folds, $n_s$ close to 0.96 and $r$ about 0.15 for a large portion of the evolution. Finally, the least value of $\chi^2$ is 12.49, which is comparable to the best results obtained for two--exponential systems in Section \ref{sec:observables}. The key difficulty remains the slow approach to the attractor of all these simple theoretical models, which is clearly at odds with the actual behavior of the CMB.

Let us conclude by stressing that our analysis is clearly provisional, since it rests on simple semi--analytic tools, but we have verified that it can capture several interesting aspects of these systems that show up again in detailed simulations. A more satisfactory data analysis of models for the onset of inflation is left for another work \cite{gkmns}.

\vskip 12pt
\section*{Acknowledgments}

We are very grateful to A.~Gruppuso and P.~Natoli for several stimulating discussions. The work of NK was supported in part by the JSPS KAKENHI Grant Number 26400253. AS in on sabbatical leave, supported in part by Scuola Normale Superiore and by INFN  (I.S. Stefi). The authors would like to thank the CERN Ph--Th Unit and Scuola Normale Superiore for the kind hospitality extended to them while this work was in progress.

\end{document}